%% file: main.tex
\documentclass[conference, 10pt, letterpaper]{IEEEtran}

\usepackage[top=50pt, left=54pt, right=54pt, bottom=58pt]{geometry}
\IEEEoverridecommandlockouts

\usepackage{float}
\usepackage{amsmath,amssymb,amsfonts}
\usepackage{algorithmic}
\usepackage{graphicx}
\usepackage{pgf}
\graphicspath{ {./images/}, {./images/case-studies} }
\usepackage{textcomp}
\usepackage{xcolor}
\def\BibTeX{{\rm B\kern-.05em{\sc i\kern-.025em b}\kern-.08em
    T\kern-.1667em\lower.7ex\hbox{E}\kern-.125emX}}

\usepackage{tikz}
\usetikzlibrary{patterns}

\newcommand{\mathdefault}[1][]{} 

\usepackage{capt-of}
\usepackage{cuted}

\usepackage{siunitx}

\usepackage[
  style=ieee,
  backend=biber,
  url=false,
  doi=false,
  isbn=false,
  sorting=none,
  hyperref,
]{biblatex}
\usepackage[hidelinks,draft=false]{hyperref}
\usepackage{csquotes}
\AtEveryBibitem{%
  \clearfield{note}%
}
\AtEveryBibitem{\clearfield{url}}
\AtEveryBibitem{\clearfield{number}}
\AtEveryBibitem{\clearfield{abstract}}
\AtEveryBibitem{\clearfield{pages}}
\AtEveryBibitem{\clearlist{language}}

\usepackage[capitalise,nameinlink,noabbrev]{cleveref}

\usepackage{gensymb}

\begin{document}

\title{Cloudy with a Chance of Green: Measuring the Predictability of 18,009 Traffic Lights in Hamburg}

\author{%
  \IEEEauthorblockN{%
    Daniel Jeschor\IEEEauthorrefmark{1},
    Philipp Matthes\IEEEauthorrefmark{1},
    Thomas Springer\IEEEauthorrefmark{1},
    Sebastian Pape\IEEEauthorrefmark{2},
    Sven Fröhlich\IEEEauthorrefmark{2}
  }%
  \IEEEauthorblockA{
    \IEEEauthorrefmark{1}\textit{Institute of Systems Architecture}, \textit{Chair of Distributed and Networked Systems, TU Dresden} \\
    \IEEEauthorrefmark{2}\textit{Insitute of Traffic Telematics}, \textit{Chair of Traffic Process Automation, TU Dresden} \\
    $[$ daniel.jeschor, philipp.matthes, thomas.springer, sven.froehlich, sebastian.pape $]$@tu-dresden.de
  } 
}

\maketitle

\begin{abstract}
Informing drivers about the predicted state of upcoming traffic lights is considered a key solution to reduce unneeded energy expenditure and dilemma zones at intersections. However, newer traffic lights can react to traffic demand, resulting in spontaneous switching behavior and poor predictability. To assess whether future traffic light assistance services are viable, it is crucial to understand how strongly predictability is affected by such spontaneous switching behavior. Previous studies have so far only reported percentages of adaptivity-capable traffic lights, but the actual switching behavior has not been measured. Addressing this research gap, we conduct a large-scale predictability evaluation based on 424 million recorded switching cycles over four weeks for 18,009 individual traffic lights in Hamburg. Two characteristics of predictability are studied: cycle discrepancy and wait time diversity. Results indicate that fewer traffic lights exhibit hard-to-predict switching behavior than suggested by previous work, considering a reported number of 90.7\% adaptive traffic lights in Hamburg. Contrasting previous work, we find that not all traffic lights capable of adaptiveness may necessarily exhibit low predictability. We critically review these results and derive avenues for future research.
\end{abstract}


\section{Introduction}

Cooperative intelligent transport systems are considered a key pillar of future urban development since they address a multitude of issues related to urban mobility \cite{seredynski_pathways_2023}. By allowing traffic participants and road infrastructure elements to communicate and collaborate with each other, not only the safety but also the efficiency of traffic can be improved. 

One key application is predicting the switching patterns of a traffic light as a vehicle approaches it. Such traffic light assistance services, known under the terms Green Light Optimal Speed Advisory (GLOSA), Eco-Approach and Departure (EAD), or Time-To-Green (TTG), allow vehicles to reduce unnecessary stops and energy expenditure by adjusting the personal driving speed to the current or upcoming green phase(s) \cite{mellegard_day_2020}. In addition, informing drivers about future color changes may avoid the safety issue of dilemma zones \cite{suzuki_new_2018}. These occur when the traffic light unexpectedly turns amber, leading to impulsive acceleration or abrupt braking.

To establish reliable traffic light assistance services in the real world, the switching behavior needs to be predicted as accurately as possible. Typically, predictions are generated by the intersection controller \cite{zweck_traffic_2013} or a cloud system \cite{6965983}, based on the real-time switching behavior. This switching behavior is recorded over a longer period to detect reoccurring patterns.

In addition, the traffic light's real-time status also plays a large role in accurate prediction. Installed vehicle detectors or public transport prioritization can change the operation of traffic lights in a matter of seconds, resulting in unstable switching patterns \cite{7013336, schweiger_elisatm_2011}. In consequence, the accuracy of predictions is deprived, leading to an unreliable traffic light assistance service.

To understand the importance of this problem, it must be studied how many traffic lights express unstable switching behavior. Diverse statements from previous works indicate a high percentage of adaptivity-capable traffic lights of up to 95\% in cities \cite{7013336}. However, in practice, adaptivity can follow constraints within the signal program, meaning that predictability is not necessarily affected. Other factors, such as traffic density, may have an impact as well. Thus, a direct measurement of predictability is required to determine the feasibility of traffic light assistance services. The lack of such direct measurements in previous studies constitutes a substantial knowledge gap.

In this paper, make the following contributions:
\begin{itemize}
    \item We aim to address this knowledge gap and measure the predictability of traffic lights by detecting instabilities in the switching behavior. The foundation for our research is provided by an open data platform in Hamburg that provides real-time data for thousands of traffic lights throughout the urban area. 
    \item We develop two novel metrics, cycle discrepancy and wait time diversity, that measure switching behavior patterns needed for prediction. In this way, we aim to tell whether current prediction methods are suitable for the seen patterns. 
    \item Based on four weeks of recorded data, we aim to draw much more reliable and detailed conclusions than previous work on the feasibility and prospect of traffic light assistance services.
\end{itemize}

The rest of this paper is structured as follows. In \Cref{sec:background}, we dissect specific types of traffic adaptivity and how these may impact predictability.  \Cref{sec:rw} summarizes previous studies on adaptive traffic lights. Our methods for data mining and predictability analysis are presented in \Cref{sec:methods}. Finally, our predictability analysis results are discussed in  \Cref{sec:results}, and concluded in \Cref{sec:conc}.

\section{Background}\label{sec:background}

\begin{figure}[t]
\centering 
\begin{tabular}{c}
\includegraphics[width=0.95\linewidth]{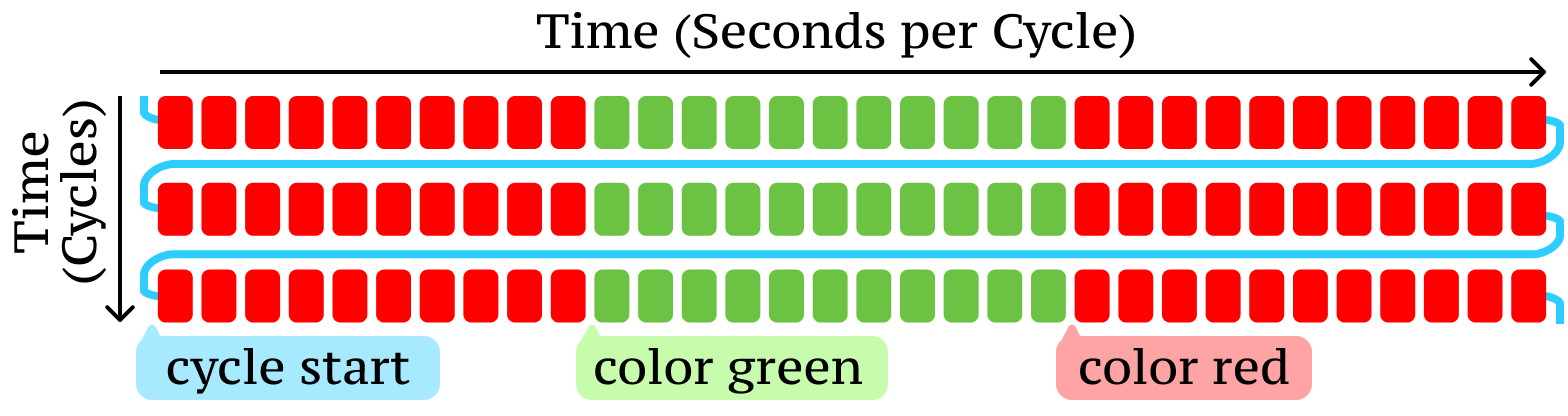} \\
\vspace{3mm}\footnotesize{(Level 1) Static / Fixed-Time} \\
\includegraphics[width=0.95\linewidth]{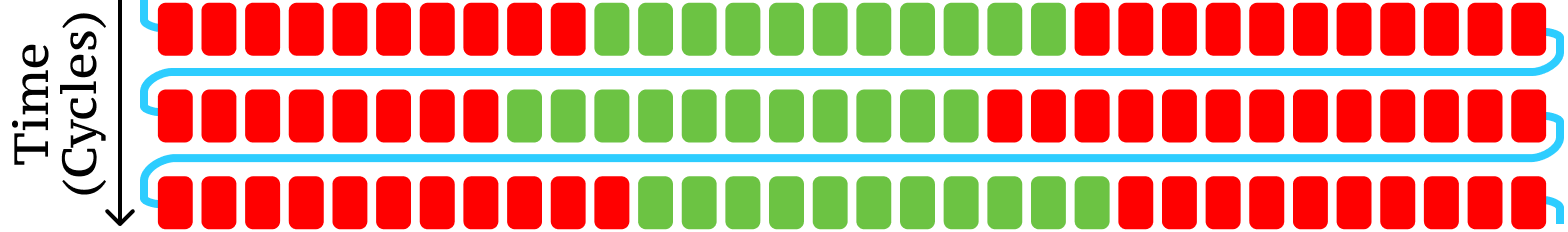} \\
\vspace{3mm}\footnotesize{(Level 2) Partially Adaptive} \\
\includegraphics[width=0.95\linewidth]{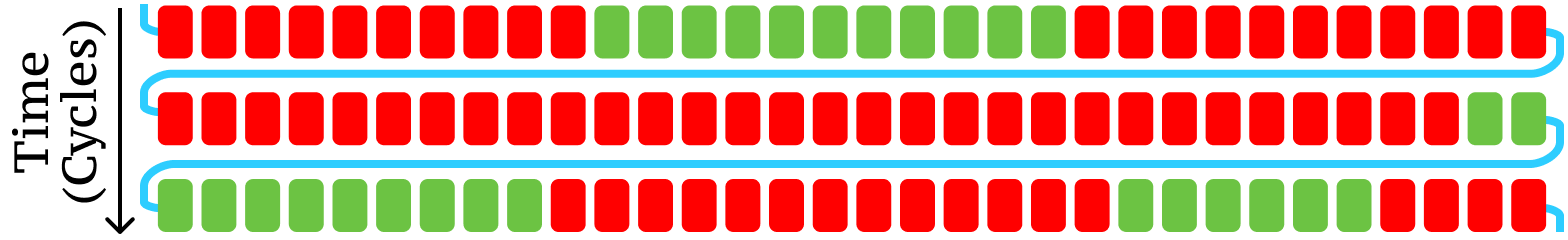} \\
\footnotesize{(Level 3) Fully Adaptive}
\end{tabular} 
\caption{Possible switching behaviors of traffic lights and the resulting patterns between cycles depending on the short-term adaptivity level.}\label{fig:case-study-cycles}
\end{figure}

The interaction of traffic lights at an intersection is orchestrated in a sequence of stages in which specific directions are given green. Each traffic light runs a program, continuously iterating between predefined states (colors) and transitions. One complete iteration through this program has been traditionally named cycle and typically has a fixed duration \cite{6965983}. Modern traffic lights still run in cycles but may be able to adapt to current conditions by updating states, stages, or programs. This adaption can occur in a long-term or short-term time frame.

\emph{Long-term adaptivity} accounts for traffic changes over the day (peak traffic times, off-peak traffic times), over the week (weekdays, Sundays, and public holidays), and over the year (normal times, holiday times, peak shopping times). Depending on current traffic or daytime, predefined programs are interchanged for longer periods of time. This kind of adaptivity is usually not a problem for traffic light prediction because it occurs at predictable times.

A much larger problem for prediction is \emph{short-term adaptivity}. This kind of adaptivity may occur spontaneously by disabling, enabling, stretching, or shortening individual states. Three levels of short-term adaptivity, as highlighted in \Cref{fig:case-study-cycles}, can be distinguished:

\begin{enumerate}
    \item \emph{Fixed-time} traffic lights always run in the same way, regardless of how the traffic situation changes.
    \item \emph{Partially adaptive} traffic lights are able to adjust their green time depending on the current traffic situation. This includes shortening and lengthening green phases or suspending and inserting individual program stages while maintaining the cycle time. If, for example, a bus is registered from the adjacent direction, the demand stage in which the bus is given clearance can be switched immediately afterward in favor of the originally planned stage.
    \item \emph{Fully adaptive} traffic lights are not based on a signal program that is changed. Instead, states are interchanged freely as demanded, while they may be constrained to maximum or minimum durations. Usually, the main direction is set to green, and the secondary direction is only enabled if there is a request.
\end{enumerate}

To determine which traffic lights are suitable for assistance services, it is crucial to distinguish which kinds of programs run on each traffic light. Based on an inquiry to Hamburg's authorities as of June 22, 2023, a mere 9.3\% (161) out of 1731 intersection nodes in the city follow a fixed-time program, leaving the remaining 90.7\% (1570) with adaptive capabilities. However, this only provides limited information about how the traffic lights actually switch. Thus, a direct analysis of the switching behavior is necessary to determine how much predictability is actually affected.

\section{Related Work}\label{sec:rw} 

At how many intersections the prediction of traffic lights could be challenging has so far been mainly conveyed through percentages separating fixed from adaptive programs. According to Bodenheimer et al. (2014) \cite{7013336}, in the ten biggest German cities, 73\% of all traffic lights are fully or partially adaptive, with a percentage of 95\% adaptive traffic lights in Hamburg, +4.3\% from our current estimate. Protschky et al. (2014) \cite{6965983} come to similar findings, observing that the majority of traffic lights are traffic responsive, given a strong focus on Munich. The time-gap control scheme, in which vehicles are let through in bulks, seems to be popular among adaptive traffic lights, according to Erdmann (2013) \cite{6698230}.

The number of traffic-adaptive signals seems to grow over time, as a follow-up study by Bodenheimer et al. (2015) \cite{7325247} reported $\approx$ 75\% of all intersections in the ten biggest German cities to be adaptive, indicating a 2\% increase. Also, larger metropolitan areas seem to employ more traffic-adaptive signals, at least in Germany. Fakler et al. (2014) \cite{23452345435} report that traffic-actuated signal control comprises $\approx$ 65\% in cities with 50-100,000 inhabitants to $\approx$ 80\% in cities with more than one million inhabitants. These findings are also referenced in more recent studies, for example, by Schneegans et al. (2022) \cite{9912747} and Heckmann et al. (2023) \cite{Heckmann2023}. Outside of Germany, Cai et al. (2009) \cite{CAI2009456} and Peng et al. (2018) \cite{8319909} have reported high adoption of traffic-adaptive capabilities as well.

However, there are also contradictory statements. Olaverri-Monreal et al. (2018) \cite{olaverri2018implementation} report that most control systems in urban areas are still pre-timed, i.e., follow a fixed-time program. Coaligning with this finding, a high prevalence of fixed timing was also reported by Yusuf et al. (2021) \cite{arifin2021recent}. The contradiction may arise from regional differences, as Avatefipour and Sadry (2018) \cite{8500246} report that fixed-time traffic lights are widespread in Malaysia. In 2017, the U.S. Department of Transportation\footnote{See: \textit{\url{https://www.fhwa.dot.gov/innovation/everydaycounts/edc-1/pdf/asct_brochure.pdf}} (retrieved on January 24, 2024)} reported widespread use of adaptive signal control in the United Kingdom, Asia, and Australia, whereas, in the U.S., those are being used on less than one percent of all signalized intersections. Thus, certain regions seem to have refrained from implementing adaptive traffic lights up to this point. Hamburg stands out with one of the highest adoption rates, making it an optimal candidate for predictability analysis.

As discussed by Bodenheimer et al. (2014) \cite{7013336}, the high numbers of adaptive traffic lights seem to be a clear motivator for studies on advanced prediction methods. In their study, the authors propose a graph-based method that becomes more accurate as the green phase is approached, counteracting adaptive shifts. Recent Machine-Learning-based methods proposed by Schneegans et al. (2023) \cite{9912747} follow a similar self-adaption. While such methods represent a key focus in current research, varying waiting times between green phases are a key challenge, as they impute temporal stability by provoking frequent adjustments to the prediction. In the following, we will refer to this type of instability as \emph{wait time diversity}.

A different method was proposed by Protschky et al. (2014) \cite{6965983}. The probabilistic approach operates by stacking recorded cycles similar to \Cref{fig:case-study-cycles}, calculating the most likely signal phase and its probability of occurrence for each second. As illustrated by Otto et al. (2023) \cite{otto_framework_2023}, this method reacts to varying green phases by blurring out the boundaries between predicted states. This aspect can be seen as a key advantage, as it allows speed advisory algorithms to concentrate on certain parts of the traffic light prediction. Such an approach was demonstrated by Mahler et al. (2012) \cite{mahler_reducing_2012} and Typaldos et al. (2023) \cite{typaldos_modified_2023}. Furthermore, Protschky et al. (2014) \cite{6965983} also show that this method is capable of producing a prediction even with highly delayed real-time traffic light data. Besides these advantages, one key drawback is that this method is vulnerable to misalignment of switching behavior between the stacked cycles. In the following, we name this type of instability \emph{cycle discrepancy}.

As both probabilistic and self-adaptive methods have individual weak points, a closer analysis of the types of instability expressed by traffic lights is required. Yet, among mainly superficial and often partially verifiable information, no study can be identified that directly measures the types of instability impacting predictability. This observation provides the entry point for our study.

\section{Methods for Predictability Analysis}\label{sec:methods}

To address the described knowledge gap, we propose a direct measurement of cycle discrepancy and wait time diversity based on collected traffic light data. The idea is to measure the presence of switching patterns, as these patterns are the foundation for current prediction methods. Afterward, by observing both metrics in conjunction, we estimate how predictability evolves throughout the day and at how many intersections poor predictability can be seen.

\subsection{Data Acquisition and Preprocessing}\label{sec:methods-data}

Real-time state observations are obtained from an open, centralized traffic light data broker, where they are pooled from 10 traffic controllers for individual city areas in Hamburg. Given are the following types of observations for each individual traffic light: state (color) changes, program changes, and a timestamp whenever a new cycle is started. Additionally, we obtain and combine car, cyclist, bus, and pedestrian detector changes to estimate how varying traffic levels affect switching behavior.

Both an MQTT client and an HTTP polling script are utilized in combination over the course of 4 weeks to obtain data from all available traffic lights in the system. Afterward, the collected data is utilized for a direct analysis of switching patterns.

The following preprocessing steps are conducted to enhance data reliability and prepare recorded observations for predictability analysis. As we obtain a continuous stream of observations, we first reconstruct cycles for each individual traffic light as depicted in \Cref{fig:case-study-cycles}. 

The cycles are then stored in a weekly table with "hourly buckets" representing one hour for each weekday and each traffic light. Data recorded over multiple weeks is overlayed onto the same hourly buckets. This rasterization allows for a computationally efficient hourly analysis of predictability over time within a week. The massive number of cycles stored in the database can be queried efficiently for each hour and each traffic light, aggregating hourly instability metrics. 

Not all traffic controllers sending data to the centralized broker are 100\% reliable. Data issues also influence predictability but make our measurements of the real switching behavior unreliable. Thus, we only measure predictability on reliable data. To detect and remove errors in the obtained data, incorrect cycles are detected and pruned based on the following ruleset:

\begin{itemize}
    \item Amber must not appear longer than 6 seconds. Additionally, red-amber must not appear longer than 2 seconds.
    \item Cycles with the following transitions are disallowed: Red to amber, amber to green, amber to red-amber, green to red-amber, red-amber to red, and red-amber to amber.
    \item Cycles must not be longer than 1.5$\times$ or shorter than 0.5$\times$ the median of directly preceding and following cycles.
\end{itemize}

The first two are derived from German traffic light operation constraints. The last one was derived empirically from preliminary analyses of occurring errors due to missing messages.

Traffic lights are excluded from further consideration when more than 10\% of the total reconstructed cycles are removed, indicating overall inconsistent or erroneous data. With the remaining traffic light data, the instability metrics are calculated to determine predictability over the weekly course for each traffic light.

\subsection{Instability Metrics and Predictability}

To measure instabilities, we detect whether reoccurring patterns in the switching behavior are present. The predictability is then derived by checking if these patterns can be detected with one of the previously proposed prediction methods.

Our first proposed metric is the cycle discrepancy. This metric aims to capture second-wise differences between individual switching cycles of a traffic light. A low cycle discrepancy indicates that green phases are switched at similar times in each cycle. Given two cycles $C_1$ and $C_2$ consisting of state seconds that may have different lengths $l_1$ and $l_2$, our metric is as follows:

\begin{equation}
\begin{array}{c}
\text{\footnotesize{Cycle}}\\
\text{\footnotesize{Discrepancy}}
\end{array}
= \sum_{i=0}^{\max(l_1, l_2)-1} \left\{
\begin{array}{ll}
1 & \text{if } i \geq l_1 \text{ or } i \geq l_2 \\
1 & \text{if } C_1[i] \neq C_2[i] \\
0 & \text{otherwise}
\end{array} \right.
\end{equation}

The resulting value is given in seconds of discrepancy between two cycles. Zero seconds mean that both cycles are equal. If a value close to zero persists throughout recorded cycles, a high stability is given. If, however, a high value is encountered, a cycle-stacking approach is likely not a good option for prediction.

By calculating the median over all pairs of cycles for all hourly buckets in a week, we obtain an hourly timeline of the median cycle discrepancy for each traffic light. The median is chosen as an aggregation robust against outliers, especially too long cycles due to missing cycle start observations. 

Evaluating the impact of certain cycle discrepancies on the stability of green phases, we additionally determine the green length for each cycle. Given in seconds, we also aggregate the green lengths of all cycles in each hourly bucket using the median.

The cycle discrepancy has a strong focus on the alignment of patterns between cycles. However, there are also prediction methods that do not rely on cycle alignment, instead predicting the time it takes to switch between different phases. Thus, we calculate a second metric that captures whether there are reoccurring wait times between green phases. If the wait time between green phases is always the same, this hints a high predictability, independent of the cycle length.

For each hourly bucket, we combine the collected cycles into a continuous sequence. This step assumes continuity between recorded cycles. In case of gaps between cycles, which may be caused by the removal of erroneous cycles, we create multiple in itself continuous sequences. Then, after each green phase, we count the waited seconds until the next green phase, given in an integer value. Based on the recorded wait times, the diversity of wait times is calculated as follows:

\begin{equation}
\begin{array}{c}
\text{\footnotesize{Wait Time}} \\
\text{\footnotesize{Diversity}}
\end{array} = \frac{\text{\footnotesize{\# Different Wait Times}}}{\text{\footnotesize{\# Total Wait Times}}}
\end{equation}

\definecolor{good}{HTML}{00bb00}
\definecolor{bad}{HTML}{ff0000}
\begin{table}[b]
\centering
\def\arraystretch{1.5}
\caption{Types of predictability mapped by our two metrics.}
\label{tab:cases}
\begin{tabular}{c|c|c|}
\multicolumn{1}{c}{} & \multicolumn{2}{c}{Cycle Discrepancy} \\
\cline{2-3}
Wait Time Diversity& Low& High\\
\hline
\multicolumn{1}{|c|}{High}& \textbf{\color{good} High predictability}$^1$& \textbf{\color{bad} Low predictability} \\
\hline
\multicolumn{1}{|c|}{Low}& \textbf{\color{good} High predictability}& \textbf{\color{good} High predictability}$^2$\\
\hline
\multicolumn{3}{l}{$^1$\footnotesize{Unpredictable wait time between green phases.}} \\
\multicolumn{3}{l}{$^2$\footnotesize{Unpredictable with cycle-stacking prediction method.}}
\end{tabular}
\end{table}

The calculated value is mapped from 0\% to 100\% wait time diversity. The order of recurring patterns is not considered. However, even for random orders, if only a few different waiting times are seen, predictions using self-adaption are more likely to be correct, given the limited number of possibilities. In the opposite case, the waiting time changes frequently, provoking frequent adjustments to an adaptive prediction. If this case is detected, a self-adaptive prediction approach is likely not a good option.

One special case is when only a few green/red phases are switched throughout an hour. In this case, the wait time diversity is high, as it is highly unpredictable at which point in time the green/red phase will reoccur. Typically, this case arises with traffic lights that don't switch for most of the time, meaning that a prediction assuming a continuation of the current phase is likely accurate, resulting in high predictability. 

This case can be cross-examined through the cycle discrepancy since many continuous red cycles are detected, resulting in a low value. Thus, it is required to combine both metrics to determine a reliable indication of overall predictability.

\begin{figure}[!b]
  \vspace{-4mm}
  \resizebox{\linewidth}{!}{\input{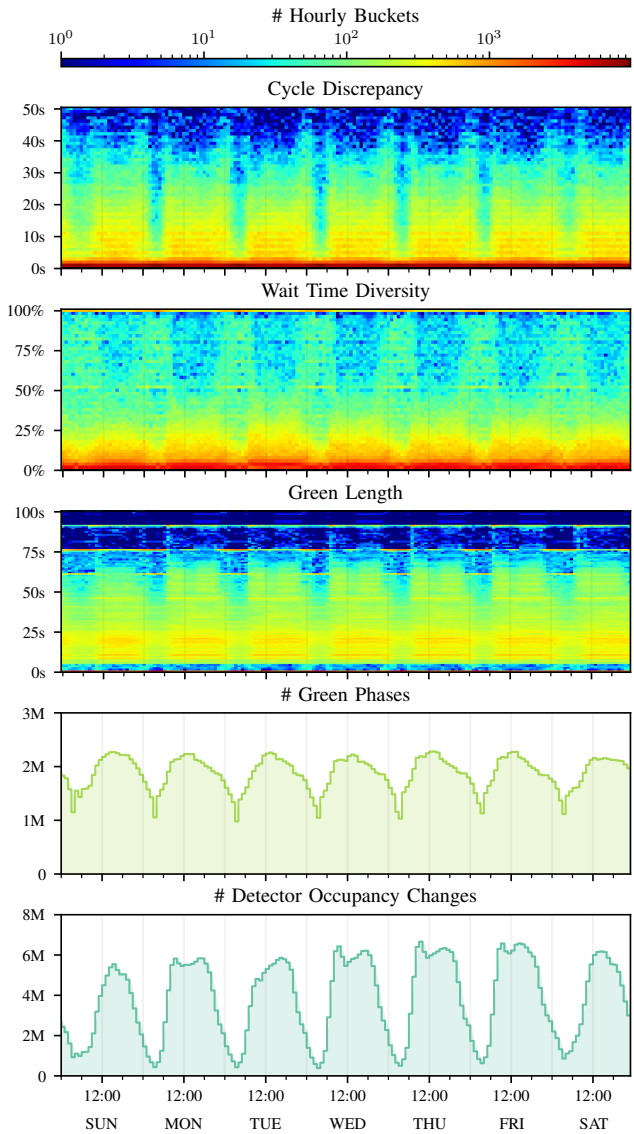}}
  \caption{Predictability for all traffic lights measured over the weekly course. Cycle discrepancy and wait time diversity values close to zero indicate high predictability. The total number of green phases and detector occupancy changes are shown as a reference for measured traffic volume. The influence on green lengths is also shown.}
  \label{fig:distance-weekly}
\end{figure}

Combining both metrics, we obtain a schema as highlighted in \Cref{tab:cases} that models predictability with current approaches. In case both instability metrics are high, the predictability can be considered generally low. Similarly, low cycle discrepancy and low wait time diversity indicate high predictability -- patterns in this area are likely similar to fixed-time programs or express minor adaptivity. In case one of both instability metrics is high, further consideration is required, as not all prediction methods may apply. For bottom-right cases, cycle-stacking prediction methods should be avoided. For top-left cases, self-adaptive prediction methods may come with the drawback of frequent prediction adjustments resulting in poor usability.

As a result, we obtain a twofold measurement of predictability that is agnostic from the internal logic of each traffic light, as it estimates predictability purely by outside observation of the switching pattern's stability.

\section{Results}\label{sec:results}

Our evaluation is split into two parts. First, we discuss how many traffic lights sent reliable data. To ensure meaningful results, only traffic lights reliably sending data are considered for our predictability analysis. Second, we explore how predictability evolves throughout the week in relation to the observed traffic levels. For each traffic light, we estimate the predictability by checking whether none, one, or both types of unstable switching behavior are expressed. 

In addition, the measured cycle-wise green length is set in relation to the cycle discrepancy to estimate whether there are predictable overlaps between switching cycles. A map is utilized to study whether unstable switching behavior is present at most or only a few intersections and if there is a relation between traffic lights connected to the same controller. Finally, we validate our results through a randomly selected sample of recorded switching patterns.

\subsection{Collected Data}

Our experiment is conducted based on a recording from 00:00 September 23, 2023, until 00:00 October 21, 2023, comprising four full weeks. For 19,844 individual traffic lights theoretically available through the centralized data broker, 18,009 traffic lights sent data. For 519 of those, in addition to the primary signal, we also have data for secondary signals such as green arrows.

During our measurement period, traffic lights have switched their colors $\approx$1.2 billion times and started a new cycle $\approx$1 billion times. Out of those, 424 million cycles could be reconstructed. Only a part of cycles is reconstructible, as sometimes, color information is missing, even though cycle start observations are available. This can be caused by errors in the data transmission. Furthermore, there are also traffic lights that still send cycle start observations even when they are turned off (e.g., during the night). In addition to the traffic light observations, we also received 642 million observations indicating occupancy changes of the traffic detectors.

The number of cycles naturally decreases during the night as traffic lights gradually go offline. In consequence, we see a decrease of up to $\approx$ 24\% in hourly buckets that contain reconstructed cycles at night.

13.8 million reconstructed cycles were flagged as erroneous and discarded, out of which 8.4 million cycles were too long or too short. Too long \mbox{(red-)}amber phases were detected at 7 million cycles. At 6 million cycles, invalid transitions between states were detected. This type of quality assurance can only be applied for traffic lights that also switch to \mbox{(red-)}amber. 34\% of traffic lights switch between green and red and are thus not suitable. The most common were traffic lights switching between green, red, and \mbox{(red-)}amber (40\%).

In 5.7 million cases, the removal of a cycle leads to a discontinuity between recorded cycles. Here, 10\% of traffic lights comprise 85\% of these cases. Thus, for a large proportion of cases, there is a continuity of cycles in the recorded history. In many cases, multiple errors were detected in one cycle. 90\% of errors are distributed over 10\% of traffic lights, indicating that a minor group of traffic lights caused most invalid cycles. At 2106 traffic lights, more than 10\% of cycles were erroneous. These traffic lights are excluded in the subsequent analyses, assuming they switch similarly to the remaining part of the traffic lights.

\begin{figure}[t]
    \centering
   \resizebox{\linewidth}{!}{\input{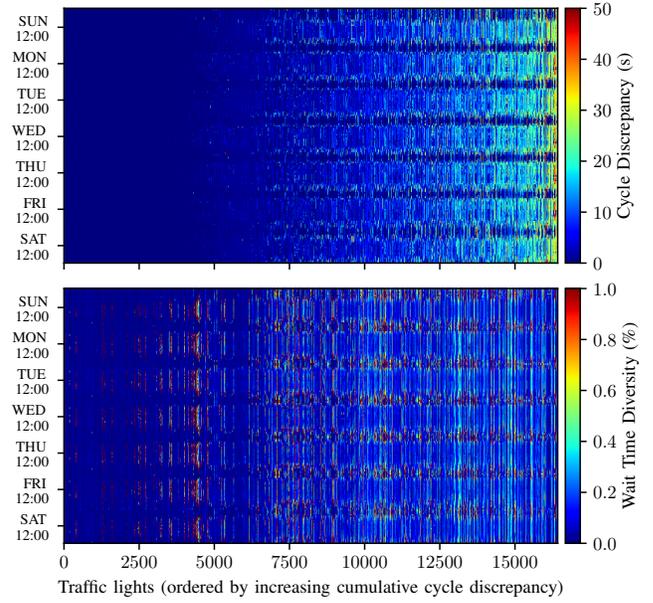}}
  \caption{Cycle discrepancy and wait time diversity, illustrating that most traffic lights only express one kind of switching instability. One vertical line, aligned between both charts, represents the weekly progression of both metrics.}
  \label{fig:week-heatmap-per-thing}
  \vspace{-3mm}
\end{figure}

\begin{figure}[t]
    \centering
  \resizebox{\linewidth}{!}{\input{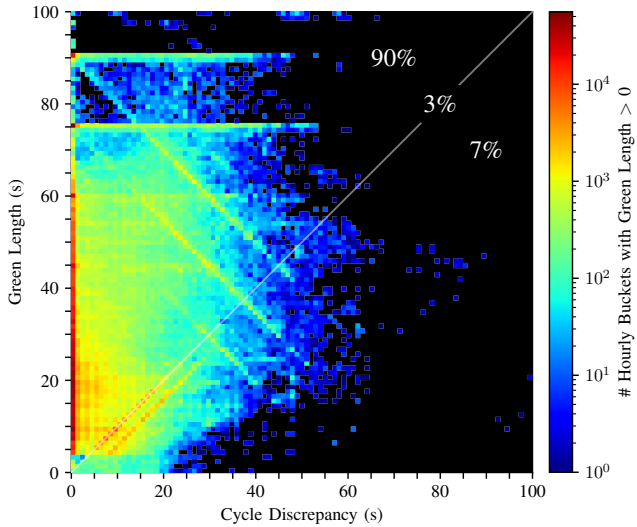}}
  \caption{Comparison between green length and cycle discrepancy, showing that the cycle discrepancy is often shorter than the green length. In 90\% of cases, there is an overlap in green phases between cycles. Line artifacts reveal specific switching patterns and constraints in the switching behavior.}
  \label{fig:cycle-dependent-instability-green-length-heatmap}
  \vspace{-3mm}
\end{figure}

\subsection{Evaluation of Traffic Light Predictability in Hamburg}

After preprocessing, the weekly progression (see \Cref{fig:distance-weekly}) of our two instability metrics shows a clear day-night rhythm, while no pronounced difference between weekends and working days is seen. During the morning traffic surge, cycle discrepancy quickly reaches a plateau with a median of 2 seconds, dropping back to 0 seconds overnight. These values are quite low, considering a 2-second cycle discrepancy only occurs at times when the median green length is between 20 seconds and 23 seconds and, thus, only comprises 10\% at maximum.

A similar plateau is seen with wait time diversity, although some traffic lights seem to express higher diversity throughout the night. This observation is likely caused by fewer switched green phases during the late evening and early morning, which increases the chance that the observed waiting times between green phases are unique. This effect may also have impacted the low nightly cycle discrepancy.

Even though cycle discrepancy and wait time diversity seem to coincide in the weekly progression, \Cref{fig:week-heatmap-per-thing} highlights that there are many cases in which only one of both types of instabilities is high. Thus, many traffic lights seem to be predictable even though they express one kind of instability pattern. Compared to these cases, the number of traffic lights expressing both instabilities seems to be low.

In numbers, 33\% of hourly buckets have a cycle discrepancy of more than 5 seconds, while 23\% of hourly buckets have a wait time diversity of more than 20\%. Hourly buckets with more than 5 seconds cycle discrepancy and more than 20\% wait time diversity only comprise 12\% of the overall distribution. Thus, an overall high switching instability seems only to be expressed by a small fraction of 90.7\% adaptivity-capable traffic lights.

As shown in \Cref{fig:cycle-dependent-instability-green-length-heatmap}, the median green length of cycles is often substantially longer than the cycle discrepancy. This type of overlap occurs in 90\% of hourly buckets with at least one green phase, often aligned to the second. When viewed in the cycle diagram, this produces "columns" of traffic light colors, as we will later see in concrete examples.

\begin{figure}[t]
   \centering
   \resizebox{\linewidth}{!}{\input{images/predictability-map.pgf}}
   \resizebox{\linewidth}{!}{\input{images/predictability-map-diversity.pgf}}
 \caption{Spatial distribution of median cycle discrepancy and wait time diversity per traffic light across Hamburg. Shown are the lanes associated with each traffic light. The zoomed-in section highlights three intersections in more detail, which express different levels of predictability.}
 \label{fig:map-capped}
 \vspace{-3mm}
\end{figure}

Specific patterns in the diagram allow a further understanding of the arrangement of green phases. First, the specified minimum green time of 5 seconds in Germany \cite{TN_libero_mab2} can be seen. For some hourly buckets, values lower than 5 seconds can also be seen, indicating cases where short green phases were often extended over a cycle. Horizontal and diagonal line artifact patterns may give us more insights into which types of constraints are utilized by traffic light programs to arrange green phases. From these patterns, we randomly extract samples to study the seen switching behavior to further investigate their composition.

The horizontal lines seem to mainly contain programs in which the green phase is active for a full cycle, occasionally interrupted by red phases. The prevalent cycle lengths of 60 seconds, 75 seconds, and 90 seconds result in the observed concentration.

Diagonal lines from top-left to bottom-right appear to come mainly from adaptive traffic lights capable of extending green phases until a maximal length. Shorter green phases can be placed more flexibly within the cycle, elevating cycle discrepancy. Increased green lengths shrink the room for placement, resulting in less cycle discrepancy. This tradeoff seems to constrain the possible adaptation, resulting in the observed linear relationship.

\Cref{fig:map-capped} shows how the two types of instability are spatially distributed across intersections. For each traffic light, the median of all hourly buckets is visualized. Based on this observation, high instability appears to be restricted to a few intersections and only rarely to single traffic lights. This result is expected, as a specific kind of traffic adaptivity is likely to affect all signals at an intersection handled by one controller. 

We cross-checked this finding by comparing the instabilities of traffic lights serving different transport modes. Based on this analysis, no noticeable differences were found between traffic lights for cars, cyclists, pedestrians, and buses -- all expressed similarly low instability. Notably, the influence of wait time diversity mainly extends to intersections distinct from those affected by cycle diversity, confirming the mutual exclusion that was seen before. In consequence, it may make sense to employ different prediction methods at different intersections.

\begin{figure}[t]
    \centering
  \resizebox{\linewidth}{!}{\input{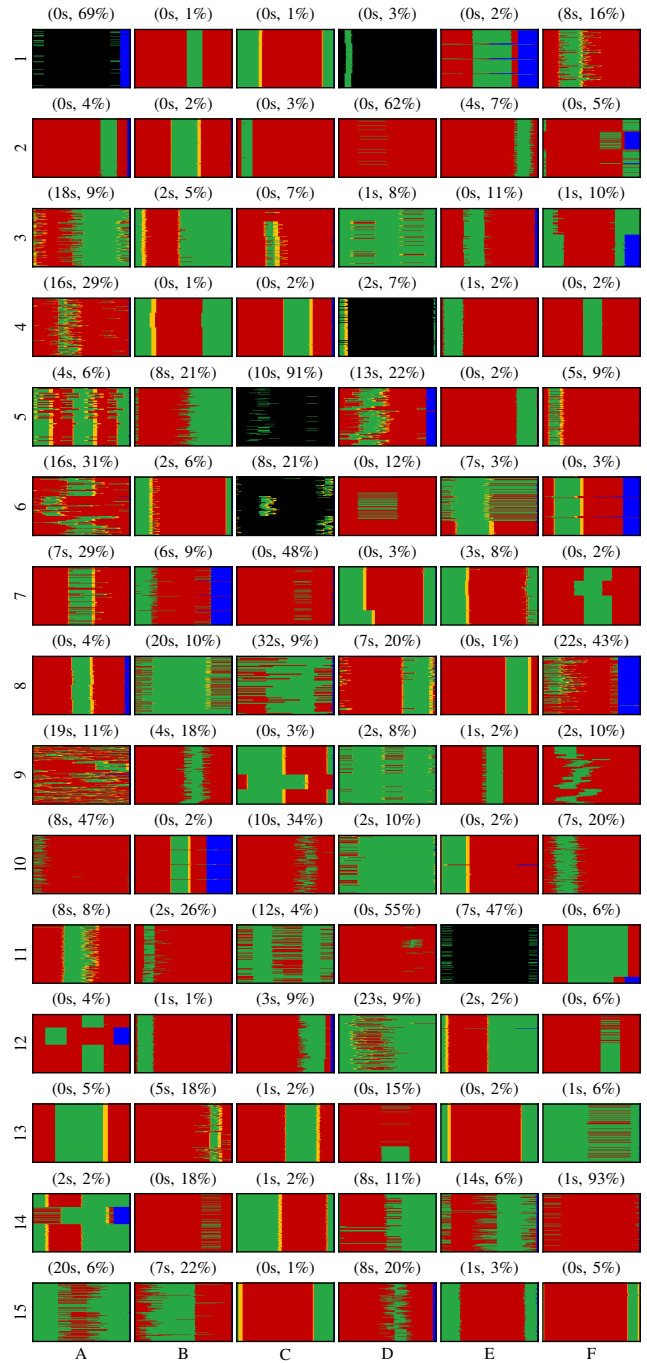}}
  \caption{Randomly selected sample of hourly buckets indicating different switching behaviors. Title format: (cycle discrepancy, wait time diversity).}
  \label{fig:predictability-case-studies}
\end{figure}

To conclude the results and cross-validate our metrics, we randomly extract hourly buckets from our database. The contained switching patterns are depicted in \ref{fig:predictability-case-studies}. 

As seen in this sample, most programs maintain a constant cycle length. In some cases, cycles switched between different lengths, as indicated by blue proportions in the diagrams. This can be due to a program change or a missed cycle start observation. Furthermore, hourly buckets comprised of the traffic light color "dark" (turned off) are also seen, which can be likely traced back to secondary signals such as green arrows.

Focusing on the type of patterns, partially adaptive programs can be seen in some examples, likely in A5, A3, F1, A11, A15, D14, or E14. Switching behaviors resembling fully adaptive traffic lights are presumably seen in A9 and F9. Some examples, such as D5 or B9, cannot be clearly distinguished. D10 and D3 show aforementioned examples in which green is the default state with occasional, fixed-length interruption by red. 

Although adaptive patterns are seen, the case study supports our main finding: most traffic lights in Hamburg express quite stable patterns -- more than expected, considering the high reported number of adaptivity-capable traffic lights. The adaption seems to be limited to a few seconds for most traffic lights, if present at all.

\subsection{Discussion}

Our results indicate that most traffic lights in Hamburg are highly predictable, suggesting only a minor part of the 90.7\% of traffic lights demonstrate their adaptivity capability to an extent that poses challenges for prediction. These results also contrast previous works, implying that low predictability cannot be assumed from a high number of adaptivity-capable traffic lights. Although highly unstable switching patterns were seen, these are restricted to a few intersection nodes. 

While switching instability increases throughout the morning traffic surge, it seems to be quickly exhausted, indicating a potential limitation of adaptivity given through traffic flow patterns or its specific implementation at the intersection controller. This is an important finding, as it indicates that traffic light assistance services may be less hindered by traffic adaptive switching behaviors than previously claimed.

However, there are also some limitations to consider. Methodologically, a key assumption is that the obtained traffic light data is trustworthy enough, supported by error detection and removal. Our hourly median aggregation assumes that instabilities, when present, are seen over 50\% of cycles for a representative measurement of predictability or statistically even out. Thus, although there may be an occasional wrong prediction, our results correspond to the chance of an accurate prediction when approaching a traffic light at a random time. 

While we consider instability measurements as an inverse proxy of predictability, it is important to encompass that each prediction method has its individual strengths and weaknesses. Where we have drawn the line between bad and good predictability, some areas for application may have stricter or less strict requirements to interpret a prediction as accurate. Finally, our analysis inherently only provides a snapshot of predictability, meaning that future measurements are required to depict a trend.

\section{Conclusion and Future Work}\label{sec:conc}

Predicting switching behavior for traffic light assistance services has been considered challenging for adaptivity-capable traffic lights. However, our study highlights that not all adaptivity-capable traffic lights may also exhibit unstable switching behavior. Thus, we find a substantial gap between theory and practice for traffic light prediction. 

Current prediction methods may not perform equally well in all situations, depending on the types of instability exhibited by a traffic light and the application scenario. Two identified weaknesses are the temporal instability of self-adaptive prediction methods and the reliance on pattern alignment between cycles with probabilistic prediction methods. Two instability metrics are proposed that, in conjunction, may help find situations in which one of both should be preferred. These may also generalize to future prediction methods.

Future work should further explore the reasons for the observed discrepancy between theory and data. Understanding whether traffic flow or the configuration of control programs imposes natural constraints on adaptivity is seen as one potential avenue, assisting in the development of traffic light assistance services.

Measurements in the future could help in understanding the long-term evolution of traffic adaptiveness. Based on related work, real-time traffic light data seems to be available in multiple cities, meaning there may be much more potential knowledge that can be extracted from large-scale data observations.

To assist such future evaluations, artifacts for our experiments can be found at \textit{\url{https://github.com/priobike/priobike-predictability-study}}.

\section{Acknowledgments}

This work is funded by the Federal Ministry for Digital and Transport (BMDV, FKZ: 16DKVM001B). Map tiles for figures are provided by Mapbox and OpenStreetMap. Traffic light data and detector data provided by the Free and Hanseatic City of Hamburg: LSBG, Data license Germany\footnote{Attribution: Version 2.0: \textit{\url{https://www.govdata.de/dl-de/by-2-0}} (retrieved on January 24, 2024) -- \textit{\url{https://metaver.de/trefferanzeige?docuuid=AB32CF78-389A-4579-9C5E-867EF31CA225}} (retrieved on January 24, 2024).}.

\renewcommand*{\bibfont}{\scriptsize}
\printbibliography
\end{document}

%% file: images/predictability-map.pgf
\begingroup%
\makeatletter%
\begin{pgfpicture}%
\pgfpathrectangle{\pgfpointorigin}{\pgfqpoint{5.627237in}{4.968148in}}%
\pgfusepath{use as bounding box, clip}%
\begin{pgfscope}%
\pgfsetbuttcap%
\pgfsetmiterjoin%
\definecolor{currentfill}{rgb}{1.000000,1.000000,1.000000}%
\pgfsetfillcolor{currentfill}%
\pgfsetlinewidth{0.000000pt}%
\definecolor{currentstroke}{rgb}{1.000000,1.000000,1.000000}%
\pgfsetstrokecolor{currentstroke}%
\pgfsetdash{}{0pt}%
\pgfpathmoveto{\pgfqpoint{0.000000in}{0.000000in}}%
\pgfpathlineto{\pgfqpoint{5.627237in}{0.000000in}}%
\pgfpathlineto{\pgfqpoint{5.627237in}{4.968148in}}%
\pgfpathlineto{\pgfqpoint{0.000000in}{4.968148in}}%
\pgfpathlineto{\pgfqpoint{0.000000in}{0.000000in}}%
\pgfpathclose%
\pgfusepath{fill}%
\end{pgfscope}%
\begin{pgfscope}%
\pgfsetbuttcap%
\pgfsetmiterjoin%
\definecolor{currentfill}{rgb}{1.000000,1.000000,1.000000}%
\pgfsetfillcolor{currentfill}%
\pgfsetlinewidth{0.000000pt}%
\definecolor{currentstroke}{rgb}{0.000000,0.000000,0.000000}%
\pgfsetstrokecolor{currentstroke}%
\pgfsetstrokeopacity{0.000000}%
\pgfsetdash{}{0pt}%
\pgfpathmoveto{\pgfqpoint{0.100600in}{0.190277in}}%
\pgfpathlineto{\pgfqpoint{4.720600in}{0.190277in}}%
\pgfpathlineto{\pgfqpoint{4.720600in}{4.810277in}}%
\pgfpathlineto{\pgfqpoint{0.100600in}{4.810277in}}%
\pgfpathlineto{\pgfqpoint{0.100600in}{0.190277in}}%
\pgfpathclose%
\pgfusepath{fill}%
\end{pgfscope}%
\begin{pgfscope}%
\pgfpathrectangle{\pgfqpoint{0.100600in}{0.190277in}}{\pgfqpoint{4.620000in}{4.620000in}}%
\pgfusepath{clip}%
\pgfsys@transformshift{0.100600in}{0.190277in}%
\pgftext[left,bottom]{\includegraphics[interpolate=true,width=4.620000in,height=4.620000in]{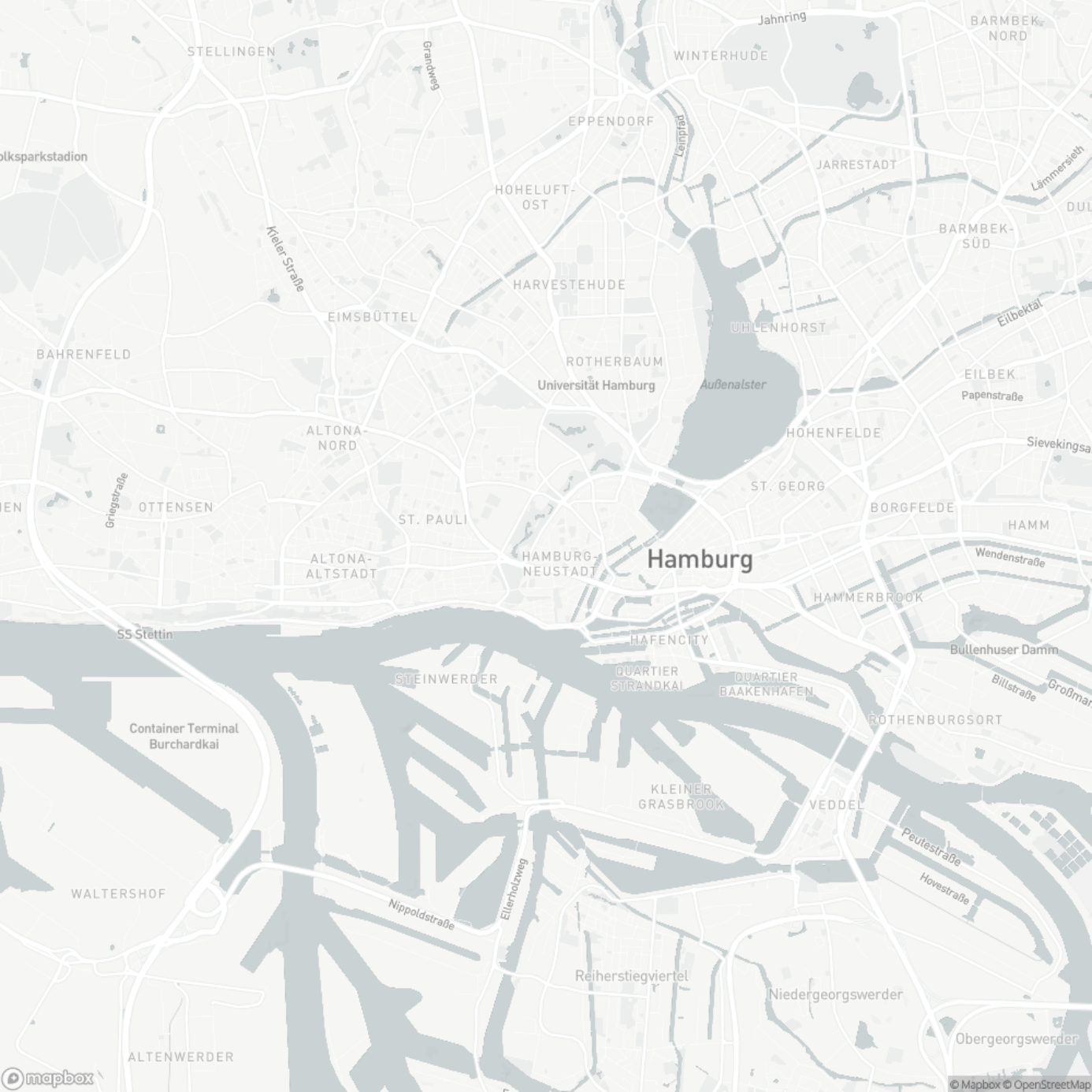}}%
\end{pgfscope}%
\begin{pgfscope}%
\pgfpathrectangle{\pgfqpoint{0.100600in}{0.190277in}}{\pgfqpoint{4.620000in}{4.620000in}}%
\pgfusepath{clip}%
\pgfsetrectcap%
\pgfsetroundjoin%
\pgfsetlinewidth{1.003750pt}%
\definecolor{currentstroke}{rgb}{0.000000,0.000000,0.000000}%
\pgfsetstrokecolor{currentstroke}%
\pgfsetdash{}{0pt}%
\pgfpathmoveto{\pgfqpoint{2.347147in}{2.622086in}}%
\pgfpathlineto{\pgfqpoint{2.347147in}{2.806906in}}%
\pgfusepath{stroke}%
\end{pgfscope}%
\begin{pgfscope}%
\pgfpathrectangle{\pgfqpoint{0.100600in}{0.190277in}}{\pgfqpoint{4.620000in}{4.620000in}}%
\pgfusepath{clip}%
\pgfsetrectcap%
\pgfsetroundjoin%
\pgfsetlinewidth{1.003750pt}%
\definecolor{currentstroke}{rgb}{0.000000,0.000000,0.000000}%
\pgfsetstrokecolor{currentstroke}%
\pgfsetdash{}{0pt}%
\pgfpathmoveto{\pgfqpoint{2.531967in}{2.622086in}}%
\pgfpathlineto{\pgfqpoint{2.531967in}{2.806906in}}%
\pgfusepath{stroke}%
\end{pgfscope}%
\begin{pgfscope}%
\pgfpathrectangle{\pgfqpoint{0.100600in}{0.190277in}}{\pgfqpoint{4.620000in}{4.620000in}}%
\pgfusepath{clip}%
\pgfsetrectcap%
\pgfsetroundjoin%
\pgfsetlinewidth{1.003750pt}%
\definecolor{currentstroke}{rgb}{0.000000,0.000000,0.000000}%
\pgfsetstrokecolor{currentstroke}%
\pgfsetdash{}{0pt}%
\pgfpathmoveto{\pgfqpoint{2.347147in}{2.622086in}}%
\pgfpathlineto{\pgfqpoint{2.531967in}{2.622086in}}%
\pgfusepath{stroke}%
\end{pgfscope}%
\begin{pgfscope}%
\pgfpathrectangle{\pgfqpoint{0.100600in}{0.190277in}}{\pgfqpoint{4.620000in}{4.620000in}}%
\pgfusepath{clip}%
\pgfsetrectcap%
\pgfsetroundjoin%
\pgfsetlinewidth{1.003750pt}%
\definecolor{currentstroke}{rgb}{0.000000,0.000000,0.000000}%
\pgfsetstrokecolor{currentstroke}%
\pgfsetdash{}{0pt}%
\pgfpathmoveto{\pgfqpoint{2.347147in}{2.806906in}}%
\pgfpathlineto{\pgfqpoint{2.531967in}{2.806906in}}%
\pgfusepath{stroke}%
\end{pgfscope}%
\begin{pgfscope}%
\pgfpathrectangle{\pgfqpoint{0.100600in}{0.190277in}}{\pgfqpoint{4.620000in}{4.620000in}}%
\pgfusepath{clip}%
\pgfsetrectcap%
\pgfsetroundjoin%
\pgfsetlinewidth{1.003750pt}%
\definecolor{currentstroke}{rgb}{0.000000,0.000000,0.000000}%
\pgfsetstrokecolor{currentstroke}%
\pgfsetdash{}{0pt}%
\pgfpathmoveto{\pgfqpoint{0.100600in}{2.038277in}}%
\pgfpathlineto{\pgfqpoint{2.347147in}{2.806906in}}%
\pgfusepath{stroke}%
\end{pgfscope}%
\begin{pgfscope}%
\pgfpathrectangle{\pgfqpoint{0.100600in}{0.190277in}}{\pgfqpoint{4.620000in}{4.620000in}}%
\pgfusepath{clip}%
\pgfsetrectcap%
\pgfsetroundjoin%
\pgfsetlinewidth{1.003750pt}%
\definecolor{currentstroke}{rgb}{0.000000,0.000000,0.000000}%
\pgfsetstrokecolor{currentstroke}%
\pgfsetdash{}{0pt}%
\pgfpathmoveto{\pgfqpoint{1.948600in}{2.038277in}}%
\pgfpathlineto{\pgfqpoint{2.531967in}{2.806906in}}%
\pgfusepath{stroke}%
\end{pgfscope}%
\begin{pgfscope}%
\pgfpathrectangle{\pgfqpoint{0.100600in}{0.190277in}}{\pgfqpoint{4.620000in}{4.620000in}}%
\pgfusepath{clip}%
\pgfsetrectcap%
\pgfsetroundjoin%
\pgfsetlinewidth{1.003750pt}%
\definecolor{currentstroke}{rgb}{0.000000,0.000000,0.000000}%
\pgfsetstrokecolor{currentstroke}%
\pgfsetdash{}{0pt}%
\pgfpathmoveto{\pgfqpoint{0.100600in}{0.190277in}}%
\pgfpathlineto{\pgfqpoint{2.347147in}{2.622086in}}%
\pgfusepath{stroke}%
\end{pgfscope}%
\begin{pgfscope}%
\pgfpathrectangle{\pgfqpoint{0.100600in}{0.190277in}}{\pgfqpoint{4.620000in}{4.620000in}}%
\pgfusepath{clip}%
\pgfsetrectcap%
\pgfsetroundjoin%
\pgfsetlinewidth{1.003750pt}%
\definecolor{currentstroke}{rgb}{0.000000,0.000000,0.000000}%
\pgfsetstrokecolor{currentstroke}%
\pgfsetdash{}{0pt}%
\pgfpathmoveto{\pgfqpoint{1.948600in}{0.190277in}}%
\pgfpathlineto{\pgfqpoint{2.531967in}{2.622086in}}%
\pgfusepath{stroke}%
\end{pgfscope}%
\begin{pgfscope}%
\pgfsys@transformshift{0.100000in}{0.191481in}%
\pgftext[left,bottom]{\includegraphics[interpolate=true,width=4.620000in,height=4.620000in]{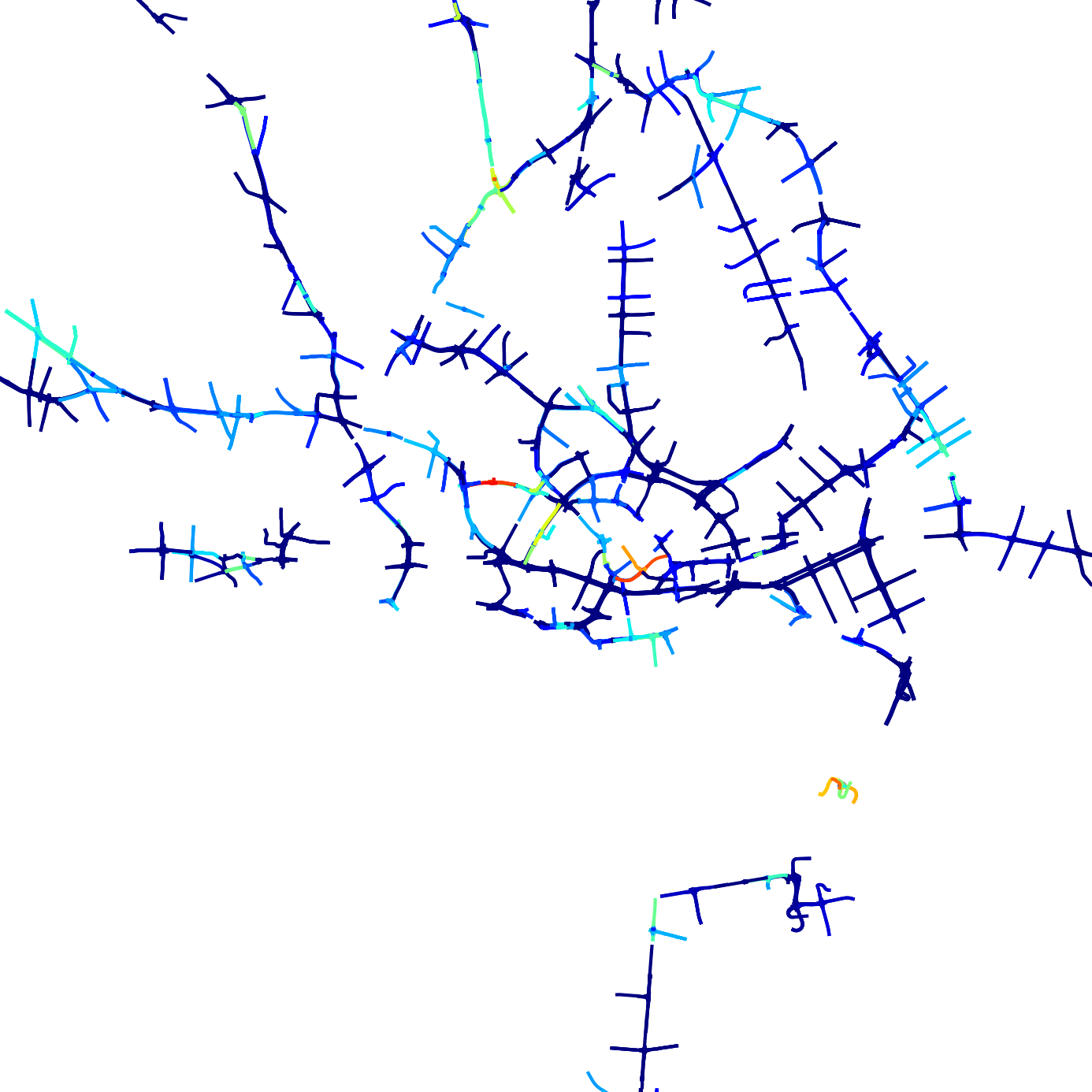}}%
\end{pgfscope}%
\begin{pgfscope}%
\pgfsetrectcap%
\pgfsetmiterjoin%
\pgfsetlinewidth{0.803000pt}%
\definecolor{currentstroke}{rgb}{0.000000,0.000000,0.000000}%
\pgfsetstrokecolor{currentstroke}%
\pgfsetdash{}{0pt}%
\pgfpathmoveto{\pgfqpoint{0.100600in}{0.190277in}}%
\pgfpathlineto{\pgfqpoint{0.100600in}{4.810277in}}%
\pgfusepath{stroke}%
\end{pgfscope}%
\begin{pgfscope}%
\pgfsetrectcap%
\pgfsetmiterjoin%
\pgfsetlinewidth{0.803000pt}%
\definecolor{currentstroke}{rgb}{0.000000,0.000000,0.000000}%
\pgfsetstrokecolor{currentstroke}%
\pgfsetdash{}{0pt}%
\pgfpathmoveto{\pgfqpoint{4.720600in}{0.190277in}}%
\pgfpathlineto{\pgfqpoint{4.720600in}{4.810277in}}%
\pgfusepath{stroke}%
\end{pgfscope}%
\begin{pgfscope}%
\pgfsetrectcap%
\pgfsetmiterjoin%
\pgfsetlinewidth{0.803000pt}%
\definecolor{currentstroke}{rgb}{0.000000,0.000000,0.000000}%
\pgfsetstrokecolor{currentstroke}%
\pgfsetdash{}{0pt}%
\pgfpathmoveto{\pgfqpoint{0.100600in}{0.190277in}}%
\pgfpathlineto{\pgfqpoint{4.720600in}{0.190277in}}%
\pgfusepath{stroke}%
\end{pgfscope}%
\begin{pgfscope}%
\pgfsetrectcap%
\pgfsetmiterjoin%
\pgfsetlinewidth{0.803000pt}%
\definecolor{currentstroke}{rgb}{0.000000,0.000000,0.000000}%
\pgfsetstrokecolor{currentstroke}%
\pgfsetdash{}{0pt}%
\pgfpathmoveto{\pgfqpoint{0.100600in}{4.810277in}}%
\pgfpathlineto{\pgfqpoint{4.720600in}{4.810277in}}%
\pgfusepath{stroke}%
\end{pgfscope}%
\begin{pgfscope}%
\pgfsetbuttcap%
\pgfsetmiterjoin%
\definecolor{currentfill}{rgb}{1.000000,1.000000,1.000000}%
\pgfsetfillcolor{currentfill}%
\pgfsetlinewidth{0.000000pt}%
\definecolor{currentstroke}{rgb}{0.000000,0.000000,0.000000}%
\pgfsetstrokecolor{currentstroke}%
\pgfsetstrokeopacity{0.000000}%
\pgfsetdash{}{0pt}%
\pgfpathmoveto{\pgfqpoint{0.100000in}{0.190277in}}%
\pgfpathlineto{\pgfqpoint{1.960000in}{0.190277in}}%
\pgfpathlineto{\pgfqpoint{1.960000in}{2.050277in}}%
\pgfpathlineto{\pgfqpoint{0.100000in}{2.050277in}}%
\pgfpathlineto{\pgfqpoint{0.100000in}{0.190277in}}%
\pgfpathclose%
\pgfusepath{fill}%
\end{pgfscope}%
\begin{pgfscope}%
\pgfpathrectangle{\pgfqpoint{0.100000in}{0.190277in}}{\pgfqpoint{1.860000in}{1.860000in}}%
\pgfusepath{clip}%
\pgfsys@transformshift{0.100000in}{0.190277in}%
\pgftext[left,bottom]{\includegraphics[interpolate=true,width=1.860000in,height=1.860000in]{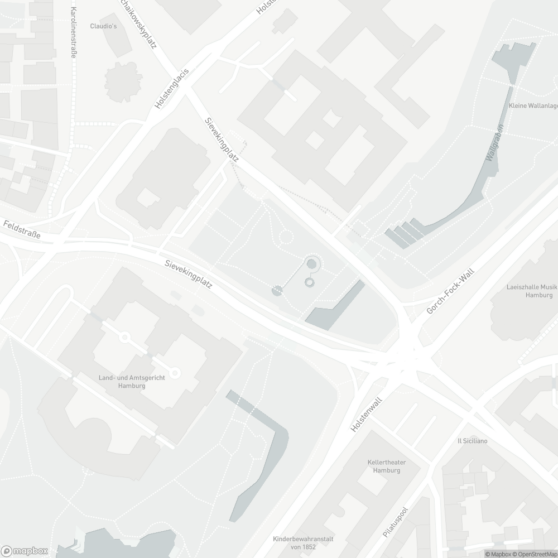}}%
\end{pgfscope}%
\begin{pgfscope}%
\pgfsys@transformshift{0.100000in}{0.191481in}%
\pgftext[left,bottom]{\includegraphics[interpolate=true,width=1.860000in,height=1.860000in]{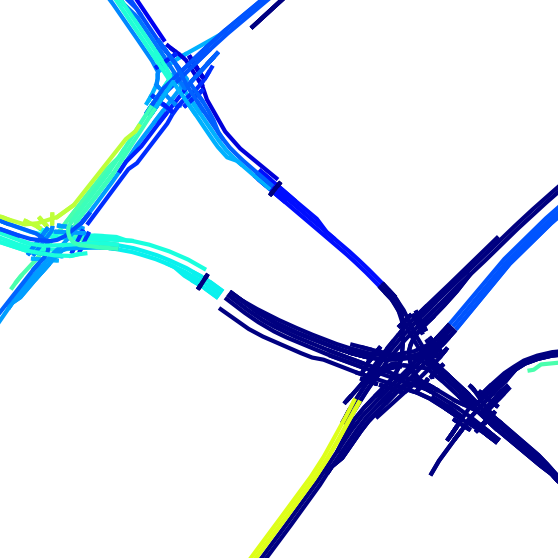}}%
\end{pgfscope}%
\begin{pgfscope}%
\pgfsetrectcap%
\pgfsetmiterjoin%
\pgfsetlinewidth{0.803000pt}%
\definecolor{currentstroke}{rgb}{0.000000,0.000000,0.000000}%
\pgfsetstrokecolor{currentstroke}%
\pgfsetdash{}{0pt}%
\pgfpathmoveto{\pgfqpoint{0.100000in}{0.190277in}}%
\pgfpathlineto{\pgfqpoint{0.100000in}{2.050277in}}%
\pgfusepath{stroke}%
\end{pgfscope}%
\begin{pgfscope}%
\pgfsetrectcap%
\pgfsetmiterjoin%
\pgfsetlinewidth{0.803000pt}%
\definecolor{currentstroke}{rgb}{0.000000,0.000000,0.000000}%
\pgfsetstrokecolor{currentstroke}%
\pgfsetdash{}{0pt}%
\pgfpathmoveto{\pgfqpoint{1.960000in}{0.190277in}}%
\pgfpathlineto{\pgfqpoint{1.960000in}{2.050277in}}%
\pgfusepath{stroke}%
\end{pgfscope}%
\begin{pgfscope}%
\pgfsetrectcap%
\pgfsetmiterjoin%
\pgfsetlinewidth{0.803000pt}%
\definecolor{currentstroke}{rgb}{0.000000,0.000000,0.000000}%
\pgfsetstrokecolor{currentstroke}%
\pgfsetdash{}{0pt}%
\pgfpathmoveto{\pgfqpoint{0.100000in}{0.190277in}}%
\pgfpathlineto{\pgfqpoint{1.960000in}{0.190277in}}%
\pgfusepath{stroke}%
\end{pgfscope}%
\begin{pgfscope}%
\pgfsetrectcap%
\pgfsetmiterjoin%
\pgfsetlinewidth{0.803000pt}%
\definecolor{currentstroke}{rgb}{0.000000,0.000000,0.000000}%
\pgfsetstrokecolor{currentstroke}%
\pgfsetdash{}{0pt}%
\pgfpathmoveto{\pgfqpoint{0.100000in}{2.050277in}}%
\pgfpathlineto{\pgfqpoint{1.960000in}{2.050277in}}%
\pgfusepath{stroke}%
\end{pgfscope}%
\begin{pgfscope}%
\pgfsetbuttcap%
\pgfsetmiterjoin%
\definecolor{currentfill}{rgb}{1.000000,1.000000,1.000000}%
\pgfsetfillcolor{currentfill}%
\pgfsetlinewidth{0.000000pt}%
\definecolor{currentstroke}{rgb}{0.000000,0.000000,0.000000}%
\pgfsetstrokecolor{currentstroke}%
\pgfsetstrokeopacity{0.000000}%
\pgfsetdash{}{0pt}%
\pgfpathmoveto{\pgfqpoint{4.813600in}{0.190277in}}%
\pgfpathlineto{\pgfqpoint{5.044600in}{0.190277in}}%
\pgfpathlineto{\pgfqpoint{5.044600in}{4.810277in}}%
\pgfpathlineto{\pgfqpoint{4.813600in}{4.810277in}}%
\pgfpathlineto{\pgfqpoint{4.813600in}{0.190277in}}%
\pgfpathclose%
\pgfusepath{fill}%
\end{pgfscope}%
\begin{pgfscope}%
\pgfsys@transformshift{4.813333in}{0.191481in}%
\pgftext[left,bottom]{\includegraphics[interpolate=true,width=0.230000in,height=4.620000in]{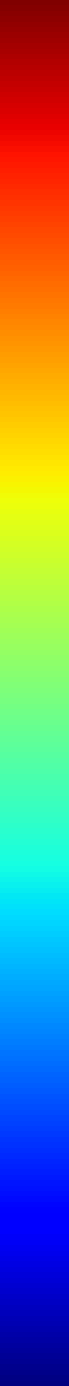}}%
\end{pgfscope}%
\begin{pgfscope}%
\pgfsetbuttcap%
\pgfsetroundjoin%
\definecolor{currentfill}{rgb}{0.000000,0.000000,0.000000}%
\pgfsetfillcolor{currentfill}%
\pgfsetlinewidth{0.803000pt}%
\definecolor{currentstroke}{rgb}{0.000000,0.000000,0.000000}%
\pgfsetstrokecolor{currentstroke}%
\pgfsetdash{}{0pt}%
\pgfsys@defobject{currentmarker}{\pgfqpoint{0.000000in}{0.000000in}}{\pgfqpoint{0.048611in}{0.000000in}}{%
\pgfpathmoveto{\pgfqpoint{0.000000in}{0.000000in}}%
\pgfpathlineto{\pgfqpoint{0.048611in}{0.000000in}}%
\pgfusepath{stroke,fill}%
}%
\begin{pgfscope}%
\pgfsys@transformshift{5.044600in}{0.190277in}%
\pgfsys@useobject{currentmarker}{}%
\end{pgfscope}%
\end{pgfscope}%
\begin{pgfscope}%
\definecolor{textcolor}{rgb}{0.000000,0.000000,0.000000}%
\pgfsetstrokecolor{textcolor}%
\pgfsetfillcolor{textcolor}%
\pgftext[x=5.141822in, y=0.132407in, left, base]{\color{textcolor}{\rmfamily\fontsize{12.000000}{14.400000}\selectfont\catcode`\^=\active\def^{\ifmmode\sp\else\^{}\fi}\catcode`\%=\active\def
\end{pgfscope}%
\begin{pgfscope}%
\pgfsetbuttcap%
\pgfsetroundjoin%
\definecolor{currentfill}{rgb}{0.000000,0.000000,0.000000}%
\pgfsetfillcolor{currentfill}%
\pgfsetlinewidth{0.803000pt}%
\definecolor{currentstroke}{rgb}{0.000000,0.000000,0.000000}%
\pgfsetstrokecolor{currentstroke}%
\pgfsetdash{}{0pt}%
\pgfsys@defobject{currentmarker}{\pgfqpoint{0.000000in}{0.000000in}}{\pgfqpoint{0.048611in}{0.000000in}}{%
\pgfpathmoveto{\pgfqpoint{0.000000in}{0.000000in}}%
\pgfpathlineto{\pgfqpoint{0.048611in}{0.000000in}}%
\pgfusepath{stroke,fill}%
}%
\begin{pgfscope}%
\pgfsys@transformshift{5.044600in}{1.114277in}%
\pgfsys@useobject{currentmarker}{}%
\end{pgfscope}%
\end{pgfscope}%
\begin{pgfscope}%
\definecolor{textcolor}{rgb}{0.000000,0.000000,0.000000}%
\pgfsetstrokecolor{textcolor}%
\pgfsetfillcolor{textcolor}%
\pgftext[x=5.141822in, y=1.056407in, left, base]{\color{textcolor}{\rmfamily\fontsize{12.000000}{14.400000}\selectfont\catcode`\^=\active\def^{\ifmmode\sp\else\^{}\fi}\catcode`\%=\active\def
\end{pgfscope}%
\begin{pgfscope}%
\pgfsetbuttcap%
\pgfsetroundjoin%
\definecolor{currentfill}{rgb}{0.000000,0.000000,0.000000}%
\pgfsetfillcolor{currentfill}%
\pgfsetlinewidth{0.803000pt}%
\definecolor{currentstroke}{rgb}{0.000000,0.000000,0.000000}%
\pgfsetstrokecolor{currentstroke}%
\pgfsetdash{}{0pt}%
\pgfsys@defobject{currentmarker}{\pgfqpoint{0.000000in}{0.000000in}}{\pgfqpoint{0.048611in}{0.000000in}}{%
\pgfpathmoveto{\pgfqpoint{0.000000in}{0.000000in}}%
\pgfpathlineto{\pgfqpoint{0.048611in}{0.000000in}}%
\pgfusepath{stroke,fill}%
}%
\begin{pgfscope}%
\pgfsys@transformshift{5.044600in}{2.038277in}%
\pgfsys@useobject{currentmarker}{}%
\end{pgfscope}%
\end{pgfscope}%
\begin{pgfscope}%
\definecolor{textcolor}{rgb}{0.000000,0.000000,0.000000}%
\pgfsetstrokecolor{textcolor}%
\pgfsetfillcolor{textcolor}%
\pgftext[x=5.141822in, y=1.980407in, left, base]{\color{textcolor}{\rmfamily\fontsize{12.000000}{14.400000}\selectfont\catcode`\^=\active\def^{\ifmmode\sp\else\^{}\fi}\catcode`\%=\active\def
\end{pgfscope}%
\begin{pgfscope}%
\pgfsetbuttcap%
\pgfsetroundjoin%
\definecolor{currentfill}{rgb}{0.000000,0.000000,0.000000}%
\pgfsetfillcolor{currentfill}%
\pgfsetlinewidth{0.803000pt}%
\definecolor{currentstroke}{rgb}{0.000000,0.000000,0.000000}%
\pgfsetstrokecolor{currentstroke}%
\pgfsetdash{}{0pt}%
\pgfsys@defobject{currentmarker}{\pgfqpoint{0.000000in}{0.000000in}}{\pgfqpoint{0.048611in}{0.000000in}}{%
\pgfpathmoveto{\pgfqpoint{0.000000in}{0.000000in}}%
\pgfpathlineto{\pgfqpoint{0.048611in}{0.000000in}}%
\pgfusepath{stroke,fill}%
}%
\begin{pgfscope}%
\pgfsys@transformshift{5.044600in}{2.962277in}%
\pgfsys@useobject{currentmarker}{}%
\end{pgfscope}%
\end{pgfscope}%
\begin{pgfscope}%
\definecolor{textcolor}{rgb}{0.000000,0.000000,0.000000}%
\pgfsetstrokecolor{textcolor}%
\pgfsetfillcolor{textcolor}%
\pgftext[x=5.141822in, y=2.904407in, left, base]{\color{textcolor}{\rmfamily\fontsize{12.000000}{14.400000}\selectfont\catcode`\^=\active\def^{\ifmmode\sp\else\^{}\fi}\catcode`\%=\active\def
\end{pgfscope}%
\begin{pgfscope}%
\pgfsetbuttcap%
\pgfsetroundjoin%
\definecolor{currentfill}{rgb}{0.000000,0.000000,0.000000}%
\pgfsetfillcolor{currentfill}%
\pgfsetlinewidth{0.803000pt}%
\definecolor{currentstroke}{rgb}{0.000000,0.000000,0.000000}%
\pgfsetstrokecolor{currentstroke}%
\pgfsetdash{}{0pt}%
\pgfsys@defobject{currentmarker}{\pgfqpoint{0.000000in}{0.000000in}}{\pgfqpoint{0.048611in}{0.000000in}}{%
\pgfpathmoveto{\pgfqpoint{0.000000in}{0.000000in}}%
\pgfpathlineto{\pgfqpoint{0.048611in}{0.000000in}}%
\pgfusepath{stroke,fill}%
}%
\begin{pgfscope}%
\pgfsys@transformshift{5.044600in}{3.886277in}%
\pgfsys@useobject{currentmarker}{}%
\end{pgfscope}%
\end{pgfscope}%
\begin{pgfscope}%
\definecolor{textcolor}{rgb}{0.000000,0.000000,0.000000}%
\pgfsetstrokecolor{textcolor}%
\pgfsetfillcolor{textcolor}%
\pgftext[x=5.141822in, y=3.828407in, left, base]{\color{textcolor}{\rmfamily\fontsize{12.000000}{14.400000}\selectfont\catcode`\^=\active\def^{\ifmmode\sp\else\^{}\fi}\catcode`\%=\active\def
\end{pgfscope}%
\begin{pgfscope}%
\pgfsetbuttcap%
\pgfsetroundjoin%
\definecolor{currentfill}{rgb}{0.000000,0.000000,0.000000}%
\pgfsetfillcolor{currentfill}%
\pgfsetlinewidth{0.803000pt}%
\definecolor{currentstroke}{rgb}{0.000000,0.000000,0.000000}%
\pgfsetstrokecolor{currentstroke}%
\pgfsetdash{}{0pt}%
\pgfsys@defobject{currentmarker}{\pgfqpoint{0.000000in}{0.000000in}}{\pgfqpoint{0.048611in}{0.000000in}}{%
\pgfpathmoveto{\pgfqpoint{0.000000in}{0.000000in}}%
\pgfpathlineto{\pgfqpoint{0.048611in}{0.000000in}}%
\pgfusepath{stroke,fill}%
}%
\begin{pgfscope}%
\pgfsys@transformshift{5.044600in}{4.810277in}%
\pgfsys@useobject{currentmarker}{}%
\end{pgfscope}%
\end{pgfscope}%
\begin{pgfscope}%
\definecolor{textcolor}{rgb}{0.000000,0.000000,0.000000}%
\pgfsetstrokecolor{textcolor}%
\pgfsetfillcolor{textcolor}%
\pgftext[x=5.141822in, y=4.752407in, left, base]{\color{textcolor}{\rmfamily\fontsize{12.000000}{14.400000}\selectfont\catcode`\^=\active\def^{\ifmmode\sp\else\^{}\fi}\catcode`\%=\active\def
\end{pgfscope}%
\begin{pgfscope}%
\definecolor{textcolor}{rgb}{0.000000,0.000000,0.000000}%
\pgfsetstrokecolor{textcolor}%
\pgfsetfillcolor{textcolor}%
\pgftext[x=5.360571in,y=2.500277in,,top,rotate=90.000000]{\color{textcolor}{\rmfamily\fontsize{12.000000}{14.400000}\selectfont\catcode`\^=\active\def^{\ifmmode\sp\else\^{}\fi}\catcode`\%=\active\def
\end{pgfscope}%
\begin{pgfscope}%
\pgfsetrectcap%
\pgfsetmiterjoin%
\pgfsetlinewidth{0.803000pt}%
\definecolor{currentstroke}{rgb}{0.000000,0.000000,0.000000}%
\pgfsetstrokecolor{currentstroke}%
\pgfsetdash{}{0pt}%
\pgfpathmoveto{\pgfqpoint{4.813600in}{0.190277in}}%
\pgfpathlineto{\pgfqpoint{4.929100in}{0.190277in}}%
\pgfpathlineto{\pgfqpoint{5.044600in}{0.190277in}}%
\pgfpathlineto{\pgfqpoint{5.044600in}{4.810277in}}%
\pgfpathlineto{\pgfqpoint{4.929100in}{4.810277in}}%
\pgfpathlineto{\pgfqpoint{4.813600in}{4.810277in}}%
\pgfpathlineto{\pgfqpoint{4.813600in}{0.190277in}}%
\pgfpathclose%
\pgfusepath{stroke}%
\end{pgfscope}%
\end{pgfpicture}%
\makeatother%
\endgroup%

%% file: images/predictability-map-diversity.pgf
\begingroup%
\makeatletter%
\begin{pgfpicture}%
\pgfpathrectangle{\pgfpointorigin}{\pgfqpoint{5.708834in}{4.968148in}}%
\pgfusepath{use as bounding box, clip}%
\begin{pgfscope}%
\pgfsetbuttcap%
\pgfsetmiterjoin%
\definecolor{currentfill}{rgb}{1.000000,1.000000,1.000000}%
\pgfsetfillcolor{currentfill}%
\pgfsetlinewidth{0.000000pt}%
\definecolor{currentstroke}{rgb}{1.000000,1.000000,1.000000}%
\pgfsetstrokecolor{currentstroke}%
\pgfsetdash{}{0pt}%
\pgfpathmoveto{\pgfqpoint{0.000000in}{0.000000in}}%
\pgfpathlineto{\pgfqpoint{5.708834in}{0.000000in}}%
\pgfpathlineto{\pgfqpoint{5.708834in}{4.968148in}}%
\pgfpathlineto{\pgfqpoint{0.000000in}{4.968148in}}%
\pgfpathlineto{\pgfqpoint{0.000000in}{0.000000in}}%
\pgfpathclose%
\pgfusepath{fill}%
\end{pgfscope}%
\begin{pgfscope}%
\pgfsetbuttcap%
\pgfsetmiterjoin%
\definecolor{currentfill}{rgb}{1.000000,1.000000,1.000000}%
\pgfsetfillcolor{currentfill}%
\pgfsetlinewidth{0.000000pt}%
\definecolor{currentstroke}{rgb}{0.000000,0.000000,0.000000}%
\pgfsetstrokecolor{currentstroke}%
\pgfsetstrokeopacity{0.000000}%
\pgfsetdash{}{0pt}%
\pgfpathmoveto{\pgfqpoint{0.100600in}{0.190277in}}%
\pgfpathlineto{\pgfqpoint{4.720600in}{0.190277in}}%
\pgfpathlineto{\pgfqpoint{4.720600in}{4.810277in}}%
\pgfpathlineto{\pgfqpoint{0.100600in}{4.810277in}}%
\pgfpathlineto{\pgfqpoint{0.100600in}{0.190277in}}%
\pgfpathclose%
\pgfusepath{fill}%
\end{pgfscope}%
\begin{pgfscope}%
\pgfpathrectangle{\pgfqpoint{0.100600in}{0.190277in}}{\pgfqpoint{4.620000in}{4.620000in}}%
\pgfusepath{clip}%
\pgfsys@transformshift{0.100600in}{0.190277in}%
\pgftext[left,bottom]{\includegraphics[interpolate=true,width=4.620000in,height=4.620000in]{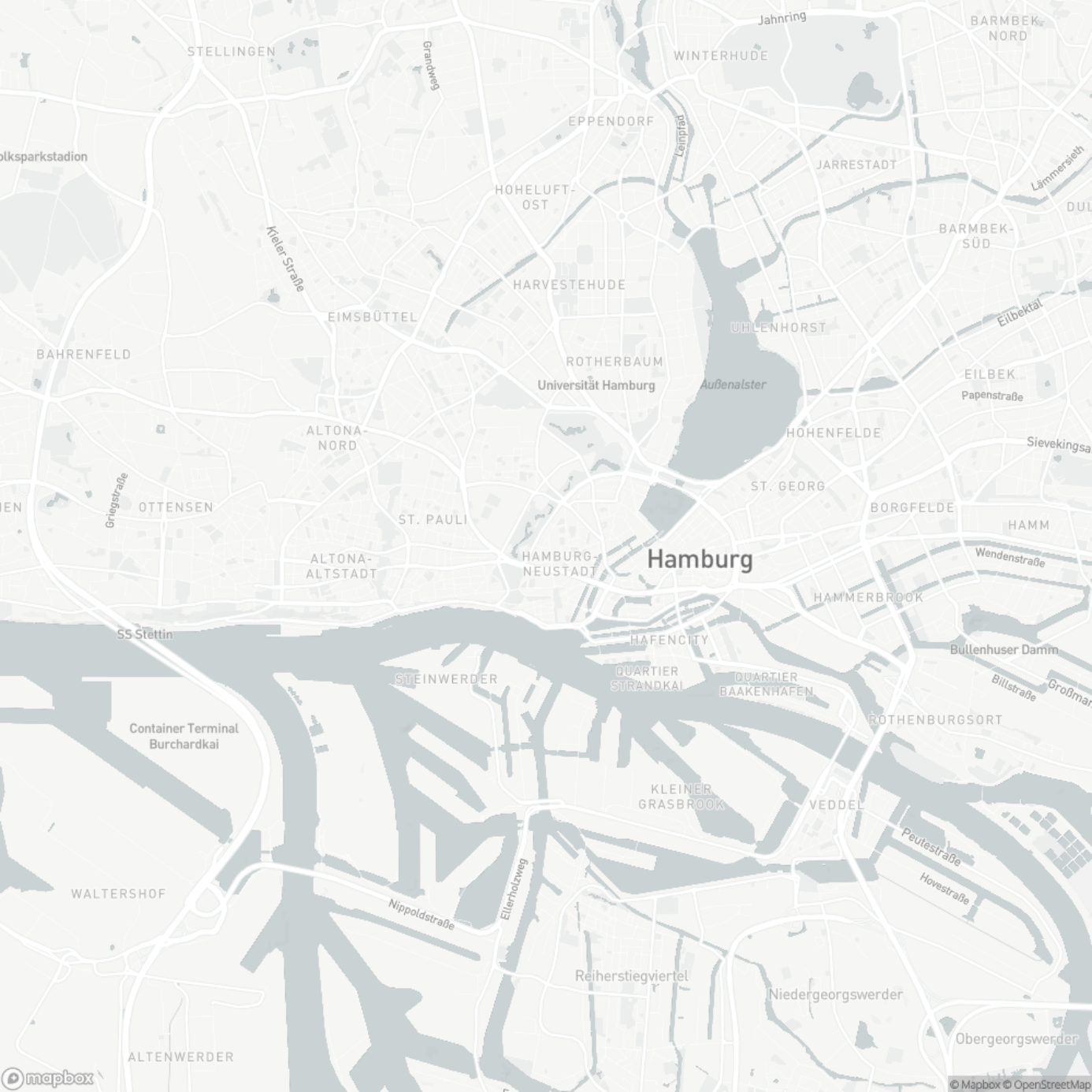}}%
\end{pgfscope}%
\begin{pgfscope}%
\pgfpathrectangle{\pgfqpoint{0.100600in}{0.190277in}}{\pgfqpoint{4.620000in}{4.620000in}}%
\pgfusepath{clip}%
\pgfsetrectcap%
\pgfsetroundjoin%
\pgfsetlinewidth{1.003750pt}%
\definecolor{currentstroke}{rgb}{0.000000,0.000000,0.000000}%
\pgfsetstrokecolor{currentstroke}%
\pgfsetdash{}{0pt}%
\pgfpathmoveto{\pgfqpoint{2.347147in}{2.622086in}}%
\pgfpathlineto{\pgfqpoint{2.347147in}{2.806906in}}%
\pgfusepath{stroke}%
\end{pgfscope}%
\begin{pgfscope}%
\pgfpathrectangle{\pgfqpoint{0.100600in}{0.190277in}}{\pgfqpoint{4.620000in}{4.620000in}}%
\pgfusepath{clip}%
\pgfsetrectcap%
\pgfsetroundjoin%
\pgfsetlinewidth{1.003750pt}%
\definecolor{currentstroke}{rgb}{0.000000,0.000000,0.000000}%
\pgfsetstrokecolor{currentstroke}%
\pgfsetdash{}{0pt}%
\pgfpathmoveto{\pgfqpoint{2.531967in}{2.622086in}}%
\pgfpathlineto{\pgfqpoint{2.531967in}{2.806906in}}%
\pgfusepath{stroke}%
\end{pgfscope}%
\begin{pgfscope}%
\pgfpathrectangle{\pgfqpoint{0.100600in}{0.190277in}}{\pgfqpoint{4.620000in}{4.620000in}}%
\pgfusepath{clip}%
\pgfsetrectcap%
\pgfsetroundjoin%
\pgfsetlinewidth{1.003750pt}%
\definecolor{currentstroke}{rgb}{0.000000,0.000000,0.000000}%
\pgfsetstrokecolor{currentstroke}%
\pgfsetdash{}{0pt}%
\pgfpathmoveto{\pgfqpoint{2.347147in}{2.622086in}}%
\pgfpathlineto{\pgfqpoint{2.531967in}{2.622086in}}%
\pgfusepath{stroke}%
\end{pgfscope}%
\begin{pgfscope}%
\pgfpathrectangle{\pgfqpoint{0.100600in}{0.190277in}}{\pgfqpoint{4.620000in}{4.620000in}}%
\pgfusepath{clip}%
\pgfsetrectcap%
\pgfsetroundjoin%
\pgfsetlinewidth{1.003750pt}%
\definecolor{currentstroke}{rgb}{0.000000,0.000000,0.000000}%
\pgfsetstrokecolor{currentstroke}%
\pgfsetdash{}{0pt}%
\pgfpathmoveto{\pgfqpoint{2.347147in}{2.806906in}}%
\pgfpathlineto{\pgfqpoint{2.531967in}{2.806906in}}%
\pgfusepath{stroke}%
\end{pgfscope}%
\begin{pgfscope}%
\pgfpathrectangle{\pgfqpoint{0.100600in}{0.190277in}}{\pgfqpoint{4.620000in}{4.620000in}}%
\pgfusepath{clip}%
\pgfsetrectcap%
\pgfsetroundjoin%
\pgfsetlinewidth{1.003750pt}%
\definecolor{currentstroke}{rgb}{0.000000,0.000000,0.000000}%
\pgfsetstrokecolor{currentstroke}%
\pgfsetdash{}{0pt}%
\pgfpathmoveto{\pgfqpoint{0.100600in}{2.038277in}}%
\pgfpathlineto{\pgfqpoint{2.347147in}{2.806906in}}%
\pgfusepath{stroke}%
\end{pgfscope}%
\begin{pgfscope}%
\pgfpathrectangle{\pgfqpoint{0.100600in}{0.190277in}}{\pgfqpoint{4.620000in}{4.620000in}}%
\pgfusepath{clip}%
\pgfsetrectcap%
\pgfsetroundjoin%
\pgfsetlinewidth{1.003750pt}%
\definecolor{currentstroke}{rgb}{0.000000,0.000000,0.000000}%
\pgfsetstrokecolor{currentstroke}%
\pgfsetdash{}{0pt}%
\pgfpathmoveto{\pgfqpoint{1.948600in}{2.038277in}}%
\pgfpathlineto{\pgfqpoint{2.531967in}{2.806906in}}%
\pgfusepath{stroke}%
\end{pgfscope}%
\begin{pgfscope}%
\pgfpathrectangle{\pgfqpoint{0.100600in}{0.190277in}}{\pgfqpoint{4.620000in}{4.620000in}}%
\pgfusepath{clip}%
\pgfsetrectcap%
\pgfsetroundjoin%
\pgfsetlinewidth{1.003750pt}%
\definecolor{currentstroke}{rgb}{0.000000,0.000000,0.000000}%
\pgfsetstrokecolor{currentstroke}%
\pgfsetdash{}{0pt}%
\pgfpathmoveto{\pgfqpoint{0.100600in}{0.190277in}}%
\pgfpathlineto{\pgfqpoint{2.347147in}{2.622086in}}%
\pgfusepath{stroke}%
\end{pgfscope}%
\begin{pgfscope}%
\pgfpathrectangle{\pgfqpoint{0.100600in}{0.190277in}}{\pgfqpoint{4.620000in}{4.620000in}}%
\pgfusepath{clip}%
\pgfsetrectcap%
\pgfsetroundjoin%
\pgfsetlinewidth{1.003750pt}%
\definecolor{currentstroke}{rgb}{0.000000,0.000000,0.000000}%
\pgfsetstrokecolor{currentstroke}%
\pgfsetdash{}{0pt}%
\pgfpathmoveto{\pgfqpoint{1.948600in}{0.190277in}}%
\pgfpathlineto{\pgfqpoint{2.531967in}{2.622086in}}%
\pgfusepath{stroke}%
\end{pgfscope}%
\begin{pgfscope}%
\pgfsys@transformshift{0.100000in}{0.191481in}%
\pgftext[left,bottom]{\includegraphics[interpolate=true,width=4.620000in,height=4.620000in]{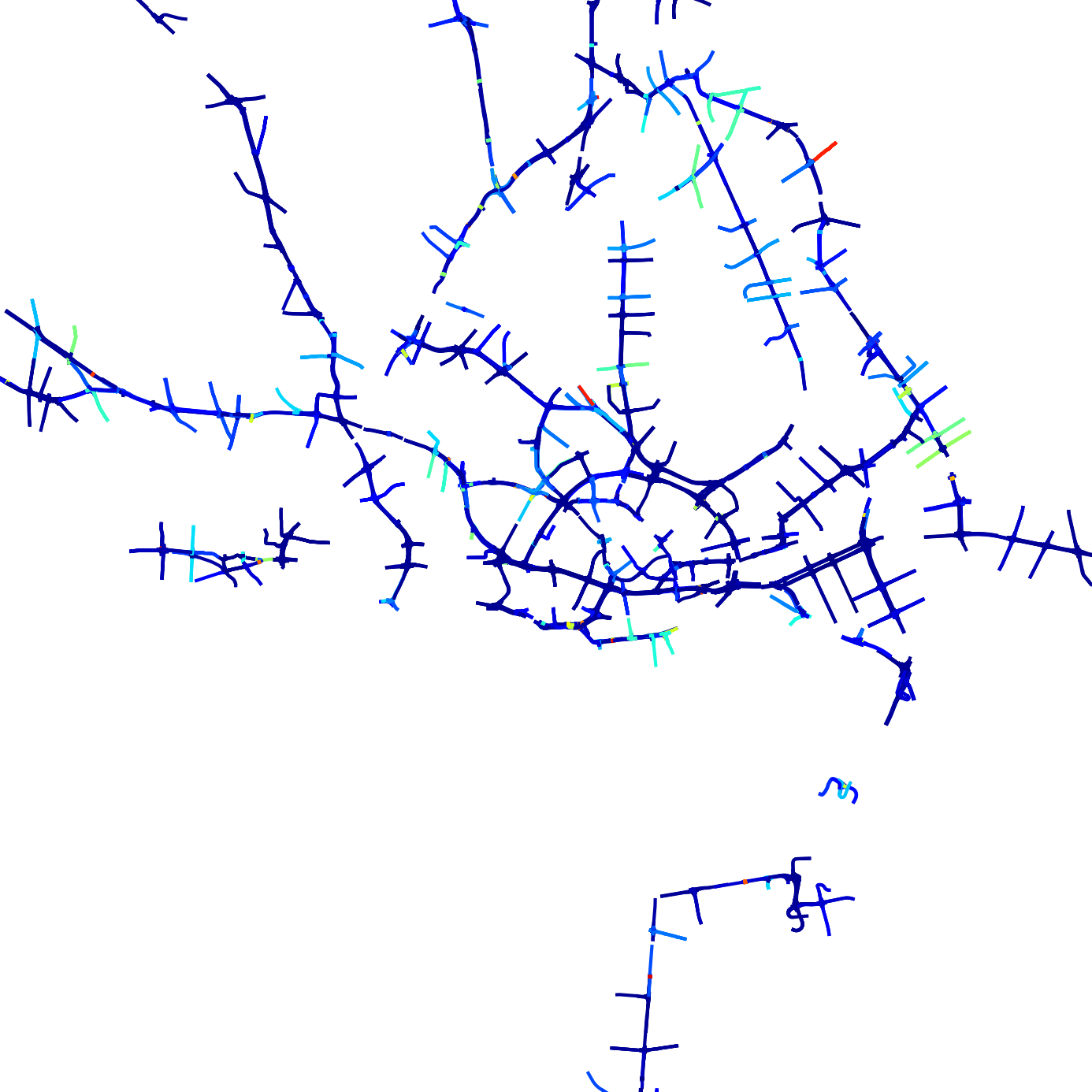}}%
\end{pgfscope}%
\begin{pgfscope}%
\pgfsetrectcap%
\pgfsetmiterjoin%
\pgfsetlinewidth{0.803000pt}%
\definecolor{currentstroke}{rgb}{0.000000,0.000000,0.000000}%
\pgfsetstrokecolor{currentstroke}%
\pgfsetdash{}{0pt}%
\pgfpathmoveto{\pgfqpoint{0.100600in}{0.190277in}}%
\pgfpathlineto{\pgfqpoint{0.100600in}{4.810277in}}%
\pgfusepath{stroke}%
\end{pgfscope}%
\begin{pgfscope}%
\pgfsetrectcap%
\pgfsetmiterjoin%
\pgfsetlinewidth{0.803000pt}%
\definecolor{currentstroke}{rgb}{0.000000,0.000000,0.000000}%
\pgfsetstrokecolor{currentstroke}%
\pgfsetdash{}{0pt}%
\pgfpathmoveto{\pgfqpoint{4.720600in}{0.190277in}}%
\pgfpathlineto{\pgfqpoint{4.720600in}{4.810277in}}%
\pgfusepath{stroke}%
\end{pgfscope}%
\begin{pgfscope}%
\pgfsetrectcap%
\pgfsetmiterjoin%
\pgfsetlinewidth{0.803000pt}%
\definecolor{currentstroke}{rgb}{0.000000,0.000000,0.000000}%
\pgfsetstrokecolor{currentstroke}%
\pgfsetdash{}{0pt}%
\pgfpathmoveto{\pgfqpoint{0.100600in}{0.190277in}}%
\pgfpathlineto{\pgfqpoint{4.720600in}{0.190277in}}%
\pgfusepath{stroke}%
\end{pgfscope}%
\begin{pgfscope}%
\pgfsetrectcap%
\pgfsetmiterjoin%
\pgfsetlinewidth{0.803000pt}%
\definecolor{currentstroke}{rgb}{0.000000,0.000000,0.000000}%
\pgfsetstrokecolor{currentstroke}%
\pgfsetdash{}{0pt}%
\pgfpathmoveto{\pgfqpoint{0.100600in}{4.810277in}}%
\pgfpathlineto{\pgfqpoint{4.720600in}{4.810277in}}%
\pgfusepath{stroke}%
\end{pgfscope}%
\begin{pgfscope}%
\pgfsetbuttcap%
\pgfsetmiterjoin%
\definecolor{currentfill}{rgb}{1.000000,1.000000,1.000000}%
\pgfsetfillcolor{currentfill}%
\pgfsetlinewidth{0.000000pt}%
\definecolor{currentstroke}{rgb}{0.000000,0.000000,0.000000}%
\pgfsetstrokecolor{currentstroke}%
\pgfsetstrokeopacity{0.000000}%
\pgfsetdash{}{0pt}%
\pgfpathmoveto{\pgfqpoint{0.100000in}{0.190277in}}%
\pgfpathlineto{\pgfqpoint{1.960000in}{0.190277in}}%
\pgfpathlineto{\pgfqpoint{1.960000in}{2.050277in}}%
\pgfpathlineto{\pgfqpoint{0.100000in}{2.050277in}}%
\pgfpathlineto{\pgfqpoint{0.100000in}{0.190277in}}%
\pgfpathclose%
\pgfusepath{fill}%
\end{pgfscope}%
\begin{pgfscope}%
\pgfpathrectangle{\pgfqpoint{0.100000in}{0.190277in}}{\pgfqpoint{1.860000in}{1.860000in}}%
\pgfusepath{clip}%
\pgfsys@transformshift{0.100000in}{0.190277in}%
\pgftext[left,bottom]{\includegraphics[interpolate=true,width=1.860000in,height=1.860000in]{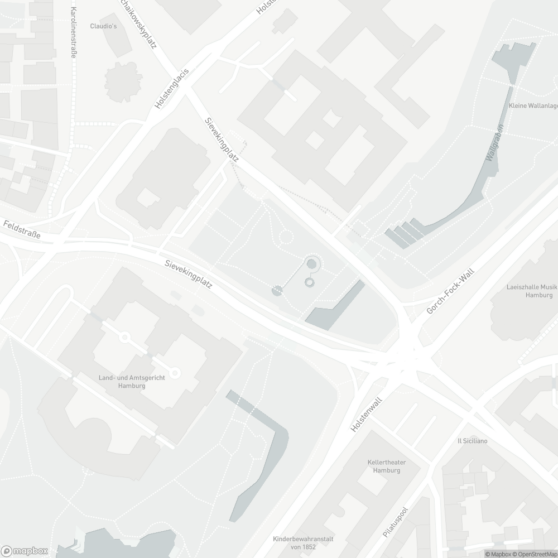}}%
\end{pgfscope}%
\begin{pgfscope}%
\pgfsys@transformshift{0.100000in}{0.191481in}%
\pgftext[left,bottom]{\includegraphics[interpolate=true,width=1.860000in,height=1.860000in]{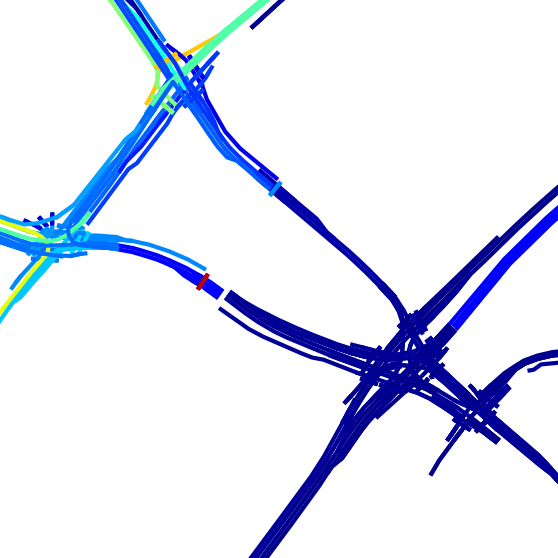}}%
\end{pgfscope}%
\begin{pgfscope}%
\pgfsetrectcap%
\pgfsetmiterjoin%
\pgfsetlinewidth{0.803000pt}%
\definecolor{currentstroke}{rgb}{0.000000,0.000000,0.000000}%
\pgfsetstrokecolor{currentstroke}%
\pgfsetdash{}{0pt}%
\pgfpathmoveto{\pgfqpoint{0.100000in}{0.190277in}}%
\pgfpathlineto{\pgfqpoint{0.100000in}{2.050277in}}%
\pgfusepath{stroke}%
\end{pgfscope}%
\begin{pgfscope}%
\pgfsetrectcap%
\pgfsetmiterjoin%
\pgfsetlinewidth{0.803000pt}%
\definecolor{currentstroke}{rgb}{0.000000,0.000000,0.000000}%
\pgfsetstrokecolor{currentstroke}%
\pgfsetdash{}{0pt}%
\pgfpathmoveto{\pgfqpoint{1.960000in}{0.190277in}}%
\pgfpathlineto{\pgfqpoint{1.960000in}{2.050277in}}%
\pgfusepath{stroke}%
\end{pgfscope}%
\begin{pgfscope}%
\pgfsetrectcap%
\pgfsetmiterjoin%
\pgfsetlinewidth{0.803000pt}%
\definecolor{currentstroke}{rgb}{0.000000,0.000000,0.000000}%
\pgfsetstrokecolor{currentstroke}%
\pgfsetdash{}{0pt}%
\pgfpathmoveto{\pgfqpoint{0.100000in}{0.190277in}}%
\pgfpathlineto{\pgfqpoint{1.960000in}{0.190277in}}%
\pgfusepath{stroke}%
\end{pgfscope}%
\begin{pgfscope}%
\pgfsetrectcap%
\pgfsetmiterjoin%
\pgfsetlinewidth{0.803000pt}%
\definecolor{currentstroke}{rgb}{0.000000,0.000000,0.000000}%
\pgfsetstrokecolor{currentstroke}%
\pgfsetdash{}{0pt}%
\pgfpathmoveto{\pgfqpoint{0.100000in}{2.050277in}}%
\pgfpathlineto{\pgfqpoint{1.960000in}{2.050277in}}%
\pgfusepath{stroke}%
\end{pgfscope}%
\begin{pgfscope}%
\pgfsetbuttcap%
\pgfsetmiterjoin%
\definecolor{currentfill}{rgb}{1.000000,1.000000,1.000000}%
\pgfsetfillcolor{currentfill}%
\pgfsetlinewidth{0.000000pt}%
\definecolor{currentstroke}{rgb}{0.000000,0.000000,0.000000}%
\pgfsetstrokecolor{currentstroke}%
\pgfsetstrokeopacity{0.000000}%
\pgfsetdash{}{0pt}%
\pgfpathmoveto{\pgfqpoint{4.813600in}{0.190277in}}%
\pgfpathlineto{\pgfqpoint{5.044600in}{0.190277in}}%
\pgfpathlineto{\pgfqpoint{5.044600in}{4.810277in}}%
\pgfpathlineto{\pgfqpoint{4.813600in}{4.810277in}}%
\pgfpathlineto{\pgfqpoint{4.813600in}{0.190277in}}%
\pgfpathclose%
\pgfusepath{fill}%
\end{pgfscope}%
\begin{pgfscope}%
\pgfsys@transformshift{4.813333in}{0.191481in}%
\pgftext[left,bottom]{\includegraphics[interpolate=true,width=0.230000in,height=4.620000in]{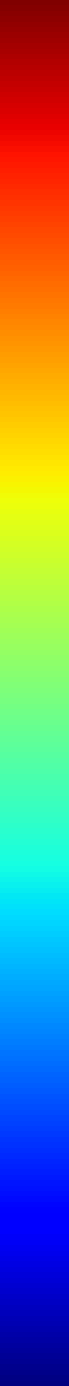}}%
\end{pgfscope}%
\begin{pgfscope}%
\pgfsetbuttcap%
\pgfsetroundjoin%
\definecolor{currentfill}{rgb}{0.000000,0.000000,0.000000}%
\pgfsetfillcolor{currentfill}%
\pgfsetlinewidth{0.803000pt}%
\definecolor{currentstroke}{rgb}{0.000000,0.000000,0.000000}%
\pgfsetstrokecolor{currentstroke}%
\pgfsetdash{}{0pt}%
\pgfsys@defobject{currentmarker}{\pgfqpoint{0.000000in}{0.000000in}}{\pgfqpoint{0.048611in}{0.000000in}}{%
\pgfpathmoveto{\pgfqpoint{0.000000in}{0.000000in}}%
\pgfpathlineto{\pgfqpoint{0.048611in}{0.000000in}}%
\pgfusepath{stroke,fill}%
}%
\begin{pgfscope}%
\pgfsys@transformshift{5.044600in}{0.190277in}%
\pgfsys@useobject{currentmarker}{}%
\end{pgfscope}%
\end{pgfscope}%
\begin{pgfscope}%
\definecolor{textcolor}{rgb}{0.000000,0.000000,0.000000}%
\pgfsetstrokecolor{textcolor}%
\pgfsetfillcolor{textcolor}%
\pgftext[x=5.141822in, y=0.132407in, left, base]{\color{textcolor}{\rmfamily\fontsize{12.000000}{14.400000}\selectfont\catcode`\^=\active\def^{\ifmmode\sp\else\^{}\fi}\catcode`\%=\active\def
\end{pgfscope}%
\begin{pgfscope}%
\pgfsetbuttcap%
\pgfsetroundjoin%
\definecolor{currentfill}{rgb}{0.000000,0.000000,0.000000}%
\pgfsetfillcolor{currentfill}%
\pgfsetlinewidth{0.803000pt}%
\definecolor{currentstroke}{rgb}{0.000000,0.000000,0.000000}%
\pgfsetstrokecolor{currentstroke}%
\pgfsetdash{}{0pt}%
\pgfsys@defobject{currentmarker}{\pgfqpoint{0.000000in}{0.000000in}}{\pgfqpoint{0.048611in}{0.000000in}}{%
\pgfpathmoveto{\pgfqpoint{0.000000in}{0.000000in}}%
\pgfpathlineto{\pgfqpoint{0.048611in}{0.000000in}}%
\pgfusepath{stroke,fill}%
}%
\begin{pgfscope}%
\pgfsys@transformshift{5.044600in}{1.114277in}%
\pgfsys@useobject{currentmarker}{}%
\end{pgfscope}%
\end{pgfscope}%
\begin{pgfscope}%
\definecolor{textcolor}{rgb}{0.000000,0.000000,0.000000}%
\pgfsetstrokecolor{textcolor}%
\pgfsetfillcolor{textcolor}%
\pgftext[x=5.141822in, y=1.056407in, left, base]{\color{textcolor}{\rmfamily\fontsize{12.000000}{14.400000}\selectfont\catcode`\^=\active\def^{\ifmmode\sp\else\^{}\fi}\catcode`\%=\active\def
\end{pgfscope}%
\begin{pgfscope}%
\pgfsetbuttcap%
\pgfsetroundjoin%
\definecolor{currentfill}{rgb}{0.000000,0.000000,0.000000}%
\pgfsetfillcolor{currentfill}%
\pgfsetlinewidth{0.803000pt}%
\definecolor{currentstroke}{rgb}{0.000000,0.000000,0.000000}%
\pgfsetstrokecolor{currentstroke}%
\pgfsetdash{}{0pt}%
\pgfsys@defobject{currentmarker}{\pgfqpoint{0.000000in}{0.000000in}}{\pgfqpoint{0.048611in}{0.000000in}}{%
\pgfpathmoveto{\pgfqpoint{0.000000in}{0.000000in}}%
\pgfpathlineto{\pgfqpoint{0.048611in}{0.000000in}}%
\pgfusepath{stroke,fill}%
}%
\begin{pgfscope}%
\pgfsys@transformshift{5.044600in}{2.038277in}%
\pgfsys@useobject{currentmarker}{}%
\end{pgfscope}%
\end{pgfscope}%
\begin{pgfscope}%
\definecolor{textcolor}{rgb}{0.000000,0.000000,0.000000}%
\pgfsetstrokecolor{textcolor}%
\pgfsetfillcolor{textcolor}%
\pgftext[x=5.141822in, y=1.980407in, left, base]{\color{textcolor}{\rmfamily\fontsize{12.000000}{14.400000}\selectfont\catcode`\^=\active\def^{\ifmmode\sp\else\^{}\fi}\catcode`\%=\active\def
\end{pgfscope}%
\begin{pgfscope}%
\pgfsetbuttcap%
\pgfsetroundjoin%
\definecolor{currentfill}{rgb}{0.000000,0.000000,0.000000}%
\pgfsetfillcolor{currentfill}%
\pgfsetlinewidth{0.803000pt}%
\definecolor{currentstroke}{rgb}{0.000000,0.000000,0.000000}%
\pgfsetstrokecolor{currentstroke}%
\pgfsetdash{}{0pt}%
\pgfsys@defobject{currentmarker}{\pgfqpoint{0.000000in}{0.000000in}}{\pgfqpoint{0.048611in}{0.000000in}}{%
\pgfpathmoveto{\pgfqpoint{0.000000in}{0.000000in}}%
\pgfpathlineto{\pgfqpoint{0.048611in}{0.000000in}}%
\pgfusepath{stroke,fill}%
}%
\begin{pgfscope}%
\pgfsys@transformshift{5.044600in}{2.962277in}%
\pgfsys@useobject{currentmarker}{}%
\end{pgfscope}%
\end{pgfscope}%
\begin{pgfscope}%
\definecolor{textcolor}{rgb}{0.000000,0.000000,0.000000}%
\pgfsetstrokecolor{textcolor}%
\pgfsetfillcolor{textcolor}%
\pgftext[x=5.141822in, y=2.904407in, left, base]{\color{textcolor}{\rmfamily\fontsize{12.000000}{14.400000}\selectfont\catcode`\^=\active\def^{\ifmmode\sp\else\^{}\fi}\catcode`\%=\active\def
\end{pgfscope}%
\begin{pgfscope}%
\pgfsetbuttcap%
\pgfsetroundjoin%
\definecolor{currentfill}{rgb}{0.000000,0.000000,0.000000}%
\pgfsetfillcolor{currentfill}%
\pgfsetlinewidth{0.803000pt}%
\definecolor{currentstroke}{rgb}{0.000000,0.000000,0.000000}%
\pgfsetstrokecolor{currentstroke}%
\pgfsetdash{}{0pt}%
\pgfsys@defobject{currentmarker}{\pgfqpoint{0.000000in}{0.000000in}}{\pgfqpoint{0.048611in}{0.000000in}}{%
\pgfpathmoveto{\pgfqpoint{0.000000in}{0.000000in}}%
\pgfpathlineto{\pgfqpoint{0.048611in}{0.000000in}}%
\pgfusepath{stroke,fill}%
}%
\begin{pgfscope}%
\pgfsys@transformshift{5.044600in}{3.886277in}%
\pgfsys@useobject{currentmarker}{}%
\end{pgfscope}%
\end{pgfscope}%
\begin{pgfscope}%
\definecolor{textcolor}{rgb}{0.000000,0.000000,0.000000}%
\pgfsetstrokecolor{textcolor}%
\pgfsetfillcolor{textcolor}%
\pgftext[x=5.141822in, y=3.828407in, left, base]{\color{textcolor}{\rmfamily\fontsize{12.000000}{14.400000}\selectfont\catcode`\^=\active\def^{\ifmmode\sp\else\^{}\fi}\catcode`\%=\active\def
\end{pgfscope}%
\begin{pgfscope}%
\pgfsetbuttcap%
\pgfsetroundjoin%
\definecolor{currentfill}{rgb}{0.000000,0.000000,0.000000}%
\pgfsetfillcolor{currentfill}%
\pgfsetlinewidth{0.803000pt}%
\definecolor{currentstroke}{rgb}{0.000000,0.000000,0.000000}%
\pgfsetstrokecolor{currentstroke}%
\pgfsetdash{}{0pt}%
\pgfsys@defobject{currentmarker}{\pgfqpoint{0.000000in}{0.000000in}}{\pgfqpoint{0.048611in}{0.000000in}}{%
\pgfpathmoveto{\pgfqpoint{0.000000in}{0.000000in}}%
\pgfpathlineto{\pgfqpoint{0.048611in}{0.000000in}}%
\pgfusepath{stroke,fill}%
}%
\begin{pgfscope}%
\pgfsys@transformshift{5.044600in}{4.810277in}%
\pgfsys@useobject{currentmarker}{}%
\end{pgfscope}%
\end{pgfscope}%
\begin{pgfscope}%
\definecolor{textcolor}{rgb}{0.000000,0.000000,0.000000}%
\pgfsetstrokecolor{textcolor}%
\pgfsetfillcolor{textcolor}%
\pgftext[x=5.141822in, y=4.752407in, left, base]{\color{textcolor}{\rmfamily\fontsize{12.000000}{14.400000}\selectfont\catcode`\^=\active\def^{\ifmmode\sp\else\^{}\fi}\catcode`\%=\active\def
\end{pgfscope}%
\begin{pgfscope}%
\definecolor{textcolor}{rgb}{0.000000,0.000000,0.000000}%
\pgfsetstrokecolor{textcolor}%
\pgfsetfillcolor{textcolor}%
\pgftext[x=5.442167in,y=2.500277in,,top,rotate=90.000000]{\color{textcolor}{\rmfamily\fontsize{12.000000}{14.400000}\selectfont\catcode`\^=\active\def^{\ifmmode\sp\else\^{}\fi}\catcode`\%=\active\def
\end{pgfscope}%
\begin{pgfscope}%
\pgfsetrectcap%
\pgfsetmiterjoin%
\pgfsetlinewidth{0.803000pt}%
\definecolor{currentstroke}{rgb}{0.000000,0.000000,0.000000}%
\pgfsetstrokecolor{currentstroke}%
\pgfsetdash{}{0pt}%
\pgfpathmoveto{\pgfqpoint{4.813600in}{0.190277in}}%
\pgfpathlineto{\pgfqpoint{4.929100in}{0.190277in}}%
\pgfpathlineto{\pgfqpoint{5.044600in}{0.190277in}}%
\pgfpathlineto{\pgfqpoint{5.044600in}{4.810277in}}%
\pgfpathlineto{\pgfqpoint{4.929100in}{4.810277in}}%
\pgfpathlineto{\pgfqpoint{4.813600in}{4.810277in}}%
\pgfpathlineto{\pgfqpoint{4.813600in}{0.190277in}}%
\pgfpathclose%
\pgfusepath{stroke}%
\end{pgfscope}%
\end{pgfpicture}%
\makeatother%
\endgroup%